\documentclass[12pt]{article}
\usepackage{epsfig}
\usepackage{graphicx}
\usepackage{a4}
\usepackage{latexsym}
\usepackage{cite}

\usepackage{color}
\usepackage{colordvi}

\usepackage{pslatex}
\usepackage[latin1]{inputenc}
\usepackage[T1]{fontenc}

\newcommand{\as}{\alpha_{\rm s}}
\newcommand{\ar}{a_{\rm s}}
\newcommand{\ra}{\rightarrow}
\newcommand{\ep}{\epsilon}
\newcommand{\sla}[1]{/\!\!\!\!\!#1}

\def\mydot{\!\cdot\!}
 
\def\z#1{{\zeta_{#1}}}
\def\ca{{C^{}_A}}
\def\cf{{C^{}_F}}
\def\nf{{n^{}_{\! f}}}

\def\a{\alpha} \def\b{\beta}     \def\e{\epsilon} 
  \def\g{\gamma}    \def\m{\mu}  
\def\n{\nu} \def\p{\pi}     
 \def\cp{{\cal P}}

\begin{document}
\setlength{\parskip}{0.2cm} \setlength{\baselineskip}{0.55cm}

\begin{titlepage}
\noindent
DESY 07-002 \hfill {\tt arXiv:0704.1740v1 [hep-ph]}\\
SFB/CPP-07-01 \\
April 2007 \\
\vspace{2.0cm}
\begin{center}
\LARGE {\bf Charged current deep-inelastic scattering } \\
\LARGE {\bf at three loops  } \\
\vspace{2.6cm}
\large
S. Moch and M. Rogal\\
\vspace{1.6cm}
\normalsize
{\it Deutsches Elektronensynchrotron DESY \\
\vspace{0.1cm}
Platanenallee 6, D--15738 Zeuthen, Germany}\\
\vfill
\large {\bf Abstract}
\vspace{-0.2cm}
\end{center}
We derive for deep-inelastic neutrino($\nu$)-proton(P) scattering in the combination $\nu P - \bar \nu P$ 
the perturbative QCD corrections to three loops 
for the charged current structure functions $F_2$, $F_L$ and $F_3$.
In leading twist approximation 
we calculate the first five odd-integer Mellin moments in the case of $F_2$ and $F_L$ 
and the first five even-integer moments in the case of $F_3$.
As a new result we obtain the coefficient functions to ${\cal O}(\alpha_s^3)$ 
while the corresponding anomalous dimensions agree with known results in the literature. 
\\
\vspace{3.0cm}
\end{titlepage}

\renewcommand{\theequation}{\thesection.\arabic{equation}}
%
% ---------------------------------------------------------------------
%
\setcounter{equation}{0}
\section{Introduction}
\label{sec:introduction}
%
% ---------------------------------------------------------------------
%

Predictions for structure functions in deep-inelastic scattering (DIS) 
including perturbative corrections in Quantum Chromodynamics (QCD)
have recently been advanced to an unprecedented level of precision 
over a wide kinematical region of Bjorken $x$ and $Q^2 = -q^2$, 
with $q$ being the momentum of the exchanged gauge boson.
The knowledge of the complete three-loop splitting functions 
for the scale evolution of unpolarized parton distributions of 
hadrons~\cite{Moch:2004pa,Vogt:2004mw} together with the second-order
coefficient functions~\cite{%
vanNeerven:1991nn,Zijlstra:1991qc,Zijlstra:1992kj,Zijlstra:1992qd,Moch:1999eb}
has completed the next-to-next-to-leading order (NNLO) approximation of massless perturbative QCD 
for the DIS structure functions $F_1$, $F_2$ and $F_3$.
In addition for electromagnetic (photon-exchange) DIS the three-loop coefficient functions 
for both $F_{\,2}$ and $F_L = F_{\,2} - 2x F_1$ have become 
available~\cite{Moch:2004xu,Vermaseren:2005qc}, the latter being 
actually required to complete the NNLO predictions,
since the leading contribution to the coefficient functions is of
first order in the strong coupling constant $\as$.

In the present article, we extend the program of calculating higher order perturbative QCD corrections 
to the structure functions of charged current DIS.
Our studies are motivated by the increasingly accurate measurements of 
neutral and charged current cross sections at HERA with a polarised beam of 
electrons and positrons~\cite{Chekanov:2003vw,Adloff:2003uh,Aktas:2005ju}.
At the same time we are also able to quantitatively improve predictions for physics 
at the front-end of a neutrino-factory, see e.g. Ref.~\cite{Mangano:2001mj}.
To be specific, we consider neutrino-proton scattering in the combination $\nu P - \bar \nu P$,
which corresponds to charged lepton-proton DIS as far as QCD corrections are concerned.
Following Refs.~\cite{Larin:1994vu,Larin:1997wd,Retey:2000nq,Moch:2001im,Blumlein:2004xt}
we compute the perturbative QCD predictions to three-loop accuracy for a number 
of fixed Mellin moments of the structure functions $F_2$, $F_L$ and $F_3$.

Within the framework of the operator product expansion (OPE), and working in Mellin space, 
$F_2^{ \nu P - \bar \nu P}$ and $F_L^{ \nu P - \bar \nu P}$ are functions of odd Mellin moments only, 
while only even moments contribute to $F_3^{\nu P -\bar \nu P}$. 
This distinction between odd and even Mellin moments is opposite to the case of the 
neutral current structure functions and also to the case of charged current structure functions 
for neutrino-proton scattering in the combination $\nu P + \bar \nu P$. 
In the latter case, the three-loop results for $F_2^{ \nu P + \bar \nu P}$ and $F_L^{ \nu P + \bar \nu P}$ 
can be directly checked in electromagnetic DIS and taken over from Refs.~\cite{Moch:2004xu,Vermaseren:2005qc}. 
Also $F_3^{\nu P + \bar \nu P}$ is known to three-loop accuracy~\cite{MVV7}
with parametrizations for the respective coefficient functions given in Ref.~\cite{Vogt:2006bt}.

Having available a limited number of fixed Mellin moments for $F_2$, $F_L$ and $F_3$ 
for both combinations of neutrino-proton scattering, i.e. $\nu P \pm \bar \nu P$ 
is a prerequisite for a subsequent complete calculation of the respective quantity 
to three loops.
With the methods of Refs.~\cite{Moch:2004pa,Vogt:2004mw,Moch:2004xu,Vermaseren:2005qc} 
at hand we have all ingredients for a future computation of the 
``all-$n$'' results in Mellin-$n$ space, or equivalently 
the complete expression in Bjorken-$x$ space after an inverse Mellin transform.
However, applying the present results we can already comment on a number of phenomenological 
issues, which we do in a companion paper~\cite{MRV1}.

The outline of this article is as follows.
In Section~\ref{sec:formalism} we briefly recall our formalism, which is based
on the optical theorem, the forward Compton amplitude and the OPE. 
Specifically we emphasize the symmetry properties of the Compton amplitude for
neutral and charged current processes and show how these select either odd or even
Mellin moments for the structure functions $F_2$, $F_L$ and $F_3$ 
depending on the process under consideration, i.e. $\nu P \pm \bar \nu P$.
In Section~\ref{sec:renormalization} we recall details of the renormalization
and give all relevant details of the calculation in Section~\ref{sec:calculation}.
Section~\ref{sec:results} contains our results for the Mellin moments 
of $F_2^{ \nu P - \bar \nu P}$, $F_L^{ \nu P - \bar \nu P}$ and $F_3^{\nu P -\bar \nu P}$ 
in numerical form. 
Finally, we conclude in Section~\ref{sec:conclusions}.
The lengthy full expressions for the new moments of the coefficient functions 
are deferred to Appendix~\ref{sec:appA} 
and some details on the OPE are given in Appendix~\ref{sec:appB}.

%
% ---------------------------------------------------------------------
%
\setcounter{equation}{0}
\section{General formalism}
\label{sec:formalism}
%
% ---------------------------------------------------------------------
%

The subject of our calculation is unpolarized inclusive deep-inelastic 
lepton-nucleon scattering,
\begin{eqnarray}
\label{eq:dis}
  l(k) \:+\: {\rm nucl}(p) \:\:\ra\:\: l^{\, \prime}(k^{\,\prime}) \:+\:  X\, ,
\end{eqnarray}
where $l(k),\, l^{\,\prime}(k^{\,\prime})$ are leptons of momenta $k$ and 
$k^{\, \prime}$, ${\rm nucl}(p)$ denotes a nucleon of momenta $p$ and 
$X$ stands for all hadronic states allowed by quantum number conservation.
In this article we are concentrating on charged current
neutrino($\nu$)-proton($P$) scattering, i.e. $\nu P$, $\bar \nu P$ via $W^{\pm}$ boson exchange.
As it is well known, the differential cross section for reaction~(\ref{eq:dis}) 
can be written as a product of leptonic $L_{\mu\nu}$ and hadronic $W_{\mu\nu}$ tensors
\begin{eqnarray}
\label{eq:diffcrosssec}
d \sigma \propto L^{\mu\nu} W_{\mu\nu}\, .
\end{eqnarray}
The leptonic tensor $L^{\mu\nu}$ for electroweak or pure electromagnetic gauge boson exchange is 
detailed in the literature, see e.g. Ref.~\cite{Yao:2006px} and will not be considered here.
The hadronic tensor in Eq.~(\ref{eq:diffcrosssec}) is given by 
\begin{eqnarray}
\label{eq:htensor}
W_{\m\n}(p,q) 
&=& \frac{1}{4\pi}
     \int d^4z\, {\rm{e}}^{{\rm{i}}q \cdot z}\langle  {{\rm nucl}, p}\vert
 J^{\dagger}_{\m}(z)J_{\n}(0)\vert {{\rm nucl},p}\rangle  
\\ 
&=& e_{\m\n}\, \frac{1}{2x}F_{L}(x,Q^2) + 
    d_{\m\n}\, \frac{1}{2x}F_{2}(x,Q^2) 
  + {\rm{i}} \e_{\m\n\a\b} \frac{p^\a q^\b}{2 p\mydot q} F_{3}(x,Q^2)\, ,\nonumber
\end{eqnarray}
where $J_{\m}$ is either an electromagnetic or a weak current and $\vert{{\rm nucl},p}\rangle$ 
is the unpolarized hadronic state with momentum $p$. 
The boson transfers momentum $q$, $Q^2=-q^2 > 0,$ and 
the Bjorken scaling variable is defined as $x=Q^2/ (2p\cdot q)$ with $0 < x \leq 1$.
The tensors $e_{\m\n}$ and $d_{\m\n}$ are given by  
\begin{eqnarray}
\label{eq:tensordef}
e_{\m\n} &=& g_{\m \n}-\frac{q_{\m} q_{\n}}{q^2} \, ,\\
d_{\m\n} &=& -g_{\m \n}-p_{\m}p_{\n}\frac{4x^2}{q^2}
              -(p_{\m}q_{\n}+p_{\n}q_{\m})\frac{2x}{q^2} \, ,
\end{eqnarray} 
and $\varepsilon_{\mu\nu\alpha\beta}$ is the totally antisymmetric tensor.
The hadron structure functions $F_{i}$, $i=L,1,2,3$ are the main subject 
of our investigations in the present paper, 
with $F_{1}$ being related to $F_{L}$ and $F_{2}$ by the Callan-Gross relation,
\begin{eqnarray}
  \label{eq:callangross}
  F_{L}(x,Q^2) = F_{2}(x,Q^2) - 2xF_{1}(x,Q^2)\, .
\end{eqnarray}
The structure function $F_{3}$ describes parity-violating effects that 
arise from vector and axial-vector interference and 
vanishes for pure electromagnetic interactions.

We are interested in the Mellin moments of the structure functions, defined as 
\begin{eqnarray}
\label{eq:mellindefF2L}
\displaystyle
F_{i}(n,Q^2) &=&
\int\limits_0^1 dx\, x^{n-2} F_{i}(x,Q^2)\, ,\quad
i = 2,L\, ;
\\
\label{eq:mellindefF3}
F_{3}(n,Q^2) &=&
\int\limits_0^1 dx\, x^{n-1} F_{3}(x,Q^2)\, .
\end{eqnarray}

The optical theorem relates the hadronic tensor in Eq.~(\ref{eq:htensor}) 
to the imaginary part of the forward scattering amplitude of 
boson-nucleon scattering, $T_{\m\n}$,
\begin{eqnarray}
\label{eq:opticaltheorem}
W_{\m \n}(p,q) &=&  \frac{1}{2\pi}\, {\rm{Im}}\, T_{\m \n}(p,q)\, .
\end{eqnarray}
The forward Compton amplitude $T_{\m\n}$ has a time-ordered product of two local currents, 
to which standard perturbation theory applies,
\begin{eqnarray}
\label{eq:forwardcompton}
T_{\m\n}(p,q) &=& {\rm{i}} \int d^4z\, {\rm{e}}^{{\rm{i}}q \cdot z}
\langle {{\rm nucl},p} \vert\,
 T \left( J^{\dagger}_{\m}(z)J_{\n}(0) \right) \vert {{\rm nucl},p}\rangle\, .
\end{eqnarray}
In the Bjorken limit, $Q^2 \rightarrow \infty$, $ x$ fixed,
the integral in Eq.~(\ref{eq:forwardcompton}) is dominated by 
the integration region near the light-cone $z^2 \sim 0$.
In this region the phase in the exponent in Eq.~(\ref{eq:forwardcompton}) 
becomes stationary for the external momentum $q$ being deep in the Euclidean region. 
Thus, we can use the OPE for a formal expansion of the current product 
in Eq.~(\ref{eq:forwardcompton}) around $z^2 \sim 0$ into a series of local composite operators 
of leading twist (see e.g. Ref.~\cite{Muta:1998vi} for details).
In terms of local operators for a time ordered product of 
the two electromagnetic or weak hadronic currents the OPE for Eq.~(\ref{eq:forwardcompton}) 
can be written in the following form
\begin{eqnarray}
\label{eq:OPE}
{\lefteqn{
 {\rm{i}} \int d^4z\, {\rm{e}}^{{\rm{i}}q \cdot z}\,
T\left( J^{\dagger}_{\mu}(z)J_{\nu}(0)\right) 
         \,=\,
2 \sum_{n,j} \biggl(\frac{2}{Q^2}\biggr)^n 
\biggl[ \biggl(g_{\mu \nu}-\frac{q_{\mu}q_{\nu}}{q^2}\biggr)
 q_{\m_1}q_{\m_2}
 C_{L,j}\biggl(n, \frac{Q^2}{\m^2},\a_s\biggr) 
}} \\
& &
-\biggl(g_{\mu \m_1}g_{\nu \m_2}q^2  -g_{\mu \m_1}q_{\nu }q_{\m_2}
   -g_{\nu \m_2}q_{\mu }q_{\m_1}   +g_{\mu \nu }q_{\m_1}q_{\m_2} \biggr)
      C_{2,j}\biggl(n,\frac{Q^2}{\m^2},\a_s\biggr)  
\nonumber\\
& &
+ {\rm{i}} \e_{\mu \nu \m_1 \beta} g^{\beta \gamma} q_{\gamma}q_{\m_2} 
C_{3,j}\biggl(n, \frac{Q^2}{\m^2},\a_s\biggr) \biggr]
q_{\m_3}...q_{\m_n} O^{j,\{\m_1,...,\m_n\}}(\m^2) 
+ {\rm{higher\,\, twists,}} 
\nonumber
\end{eqnarray}
where  $j=\alpha,{\rm{q}},{\rm{g}}$ and all quantities are assumed to be renormalized, 
$\m$ being the renormalization scale. 
Higher twist contributions are omitted in Eq.~(\ref{eq:OPE}) as they are less
singular near the light-cone $z^2 \sim 0$ and suppressed by powers of $1/Q^2$.  
Therefore, the sum over $n$ in Eq.~(\ref{eq:OPE}) extends to infinity and 
runs only over the standard set of the spin-$n$ twist-2 irreducible symmetrical and traceless operators.
In a general case three kind of operators contribute (these correspond to the index $j$ in Eq.~(\ref{eq:OPE})):
the flavor non-singlet quark operators $O^\a$, 
the flavor singlet quark operator $O^{\rm{q}}$  and the flavor singlet gluon operator $O^{\rm{g}}$.\
These are defined by,
\begin{eqnarray}
\label{eq:defoperatorns}
O^{\alpha,\{\m_1,\cdots ,\m_n\}} & = & \overline{\psi}\lambda^{\alpha}
  \gamma^{\{\m_1}D^{\m_2}\cdots D^{\m_n\}}\psi,~~\alpha=1,2,...,(n_f^2-1)\, ,\\
\label{eq:defoperatorquark}
O^{{\rm{q}},\{\m_1,\cdots ,\m_n\}} & = & \overline{\psi}
  \gamma^{\{\m_1}D^{\m_2}\cdots D^{\m_n\}}\psi, \\
\label{eq:defoperatorgluon}
O^{{\rm{g}},\{\m_1,\cdots ,\m_n\}} & = & F^{\n\{\m_1} D^{\m_2}\cdots
  D^{\m_{n-1}} F^{\m_n\} \n }\, .
\end{eqnarray}
Here, $\psi$ defines the quark operator and $F^{\m\n}$ the gluon operator.
The generators of the flavor group $SU(n_f)$ are denoted 
by $\lambda^{\alpha}$, and the covariant derivative by $D^\m$. 
It is understood that the symmetrical and traceless part 
is taken with respect to the indices in curly brackets.

The spin averaged operator matrix elements (OMEs) 
in Eqs.~(\ref{eq:defoperatorns})--(\ref{eq:defoperatorgluon}) 
sandwiched between some hadronic state are given by
\begin{eqnarray}
\label{eq:OME}
\langle {{\rm nucl},p} \vert O^{j, \{\m_1,...,\m_n\}}\vert
 {{\rm nucl},p} \rangle =p^{\{\m_1}...p^{\m_n\}}A_{{\rm{nucl}}}^j\left(n, {\m^2}\right)\, ,
\end{eqnarray}
where hadron mass effects have been neglected. 
The OMEs themselves as given in Eq.~(\ref{eq:OME}) are not calculable in perturbative QCD, 
but they can be related to the quark and anti-quark distributions of a given flavor and 
to the gluon distribution in the hadron. 

The scale evolution of the OMEs governed by anomalous dimensions is accessible 
to perturbative predictions as well as the coefficient functions $C_{i,j}$ multiplying 
the OMEs according to Eq.~(\ref{eq:OPE}).
Both the anomalous dimensions and the coefficient functions are calculable order by order 
in perturbative QCD in an expansion in the strong coupling constant $\a_s$.
In order to do so, we replace the nucleon state $\vert {\rm nucl,}p \rangle$ in
Eqs.~(\ref{eq:forwardcompton}), (\ref{eq:OPE}) by partonic states.
In complete analogy to Eq.~(\ref{eq:forwardcompton}) we define the 
forward Compton amplitude $t_{\mu\nu}$ at parton level and 
the corresponding partonic OMEs, 
\begin{eqnarray}
\label{eq:OMEpartons}
\langle {{\rm parton},p} \vert O^{j, \{\m_1,...,\m_n\}}\vert
 {{\rm parton},p} \rangle =p^{\{\m_1}...p^{\m_n\}}A_{{\rm{parton}}}^j\left(n, {\m^2}\right)\, .
\end{eqnarray}
As the OPE in Eq.~(\ref{eq:OPE}) represents an operator relation, 
we derive the following parton level expression 
\begin{eqnarray}
\label{eq:OPEpartons}
t_{\mu\nu} &\equiv&
 {\rm{i}} \int d^4z\, {\rm{e}}^{{\rm{i}}q \cdot z}
\langle {{\rm parton},p} \vert\,
 T \left( J^{\dagger}_{\m}(z)J_{\n}(0) \right) \vert {{\rm parton},p}\rangle\,
\\
&=&
2 \sum_{n,j} {\omega}^n \biggl[ e_{\m\n}\, C_{L,j}\biggl(n,\frac{Q^2}{\m^2},\a_s\biggr) 
     + d_{\m\n}\, C_{2,j}\biggl(n,\frac{Q^2}{\m^2},\a_s\biggr)  
\nonumber\\
& &\mbox{}
+ {\rm{i}} \e_{\m\n\a\b} \frac{p^\a q^\b}{p\mydot q} 
C_{3,j}\biggl(n,\frac{Q^2}{\m^2},\a_s\biggr) \biggr] A_{{\rm{parton}}}^{j}\left(n,{\m^2}\right) 
+ {\rm{higher\,\, twists}}
\, ,
\nonumber
\end{eqnarray}
which is an expansion in terms of the variable $\omega = (2 p\cdot q)/Q^2 = 1/x$  
for unphysical $\omega \rightarrow 0$  ($x \rightarrow \infty$).
The coefficients $C_{i,j}$ with $i=2,3,L$ are of course the same as the previous ones 
appearing in Eqs.~(\ref{eq:OPE}) and the scale evolution of the OMEs in Eq.~(\ref{eq:OMEpartons}) 
is controllable in perturbation theory. 

Let us in the following recall a few aspects of flavor (isospin) symmetry of
the DIS structure functions which are relevant to neutrino-nucleon scattering. 
The composite operators Eqs.~(\ref{eq:defoperatorns})--(\ref{eq:defoperatorgluon}) are either 
singlet or non-singlet operators referring to the representation of the $SU(n_{f})$ flavor group. 
In particular the non-singlet operator $O^{\alpha,\{\m_1,\cdots ,\m_n\}}$ in Eq.~(\ref{eq:defoperatorns}) 
contains the generators $\lambda^{\alpha}$ of the flavor $SU(n_{f})$.
It is well known, that for the separation of the singlet and non-singlet contributions 
to structure functions, Wilson coefficients, etc., one considers the sum and the difference 
of matrix elements for a proton $P$ and a neutron $N$, e.g.
\begin{equation}
\label{eq:isospin}
F_{i}^{eP\pm eN}\equiv  F_{i}^{eP} \pm  F_{i}^{eN}
\, ,
\quad\quad
F_{i}^{\nu P\pm \nu N}\equiv  F_{i}^{ \nu P} \pm  F_{i}^{ \nu N}
\, ,
\quad\quad\quad\quad i = 2,3,L
\, .  
\end{equation}
The combination $P+N$ singles out contributions to the singlet (isoscalar) operators 
and $P-N$ the corresponding ones to the non-singlet (isovector) operators, 
which can be seen readily as follows.
To that end, let us specialize for simplicity to the case of two flavors ($n_{f}=2$) only, 
i.e. to a $SU(2)$-isospin symmetry, 
the generalization to an arbitrary number $n_f$ of flavors being straightforward.
Then, in the $SU(2)$ example, the twist-two term $\Theta$ of the OPE consists of an  
isoscalar ($\theta_{0}$) and isovector ($\theta_{\alpha}$) part, i.e.,
\begin{equation}
\label{eq:twist-two-piece}
\Theta= \theta_{0}\, {\bf 1}+\theta_{\alpha}\lambda_{\alpha} , \quad \alpha=1,2,3 \, , 
\end{equation}
where {\bf 1} is unit matrix and $\lambda_{\alpha} = \sigma _{\alpha}/2, \, \sigma_{\alpha}$ are the usual Pauli matrices   
in fundamental representation. 
Sandwiching Eq.~(\ref{eq:twist-two-piece}) between the proton $|P \rangle$ and
neutron $|N\rangle$ states, one gets directly 
\begin{eqnarray}
\label{eq:PNexample}
  \langle P| \Theta |P \rangle + \langle N|\Theta|N\rangle  
  &=& 
  \theta_{0}\langle P|P\rangle + \theta_{\alpha}\langle P|\lambda_{\alpha}|P\rangle + 
  \theta_{0}\langle N|N\rangle + \theta_{\alpha}\langle P|\lambda_{\alpha}|P\rangle 
  \\
  &=& 
  \theta_0 + \frac{1}{2} \theta_{3} + \theta_{0}-\frac{1}{2} \theta_{3} = 2\, \theta_{0} 
  \, , 
  \nonumber
  \\
  \langle P|\Theta|P\rangle - \langle N|\Theta|N\rangle 
  &=&
  \theta_0 +\frac{1}{2} \theta_{3} - (\theta_{0}-\frac{1}{2} \theta_{3}) = \theta_{3} 
  \, .
\end{eqnarray}
Here one uses the fact that proton and neutron are eigenvectors of the $\lambda_{3}$ isospin operator 
with eigenvalues $+1/2$ and $-1/2$, respectively. 
Hence the combinations of the OMEs such as $ A_{ P \pm N }^j(n)=A_{P}^j(n) \pm A_{N}^j(n) $ 
correspond to the isoscalar part (singlet contribution) and isovector part (non-singlet contribution), respectively. 
As an upshot one can conclude that the OPE for the $P-N$ combination receives contributions 
from non-singlet quark operator $O^\alpha $ Eq.~(\ref{eq:defoperatorns}), i.e. $j=\alpha$,
in the r.h.s of Eq.~(\ref{eq:OPE}).
On the other hand, for the $P+N$ combination 
both singlet quark operator $O^{ \rm q }$ Eq.~(\ref{eq:defoperatorquark}) and
singlet gluon operator $O^{\rm g } $ 
Eq.~(\ref{eq:defoperatorgluon})  contribute in the OPE, 
i.e. the sum on the r.h.s  Eq.~(\ref{eq:OPE}) runs over $j={\rm q}, {\rm g}$. 

Since in the present article, we are considering charged current DIS 
in the combination $\nu P - \bar \nu P$ we have due to isospin symmetry 
\begin{eqnarray}
\label{eq:FnuPFnuN}
\left.
  \begin{array}{c}
    F_{i}^{\bar \nu P} = F_{i}^{ \nu N} \\[1ex]
    F_{i}^{\bar \nu N} = F_{i}^{\nu P} 
  \end{array}
\right\}
  &\Rightarrow&
  F_{i}^{ \nu P} -  F_{i}^{\bar \nu P} = 
  F_{i}^{ \nu P} -F_{i}^{\nu N} =
  F_{i}^{\bar  \nu N} -  F_{i}^{\nu N}
  \, .
\end{eqnarray}
Thus, we are entirely restricted to non-singlet quark operators for the structure functions 
$F_2^{ \nu P - \bar \nu P}$, 
$F_3^{ \nu P - \bar \nu P}$ 
and $F_L^{ \nu P - \bar \nu P}$.

\begin{figure}[ht]
  \begin{center}
    \includegraphics[width=5.0cm]{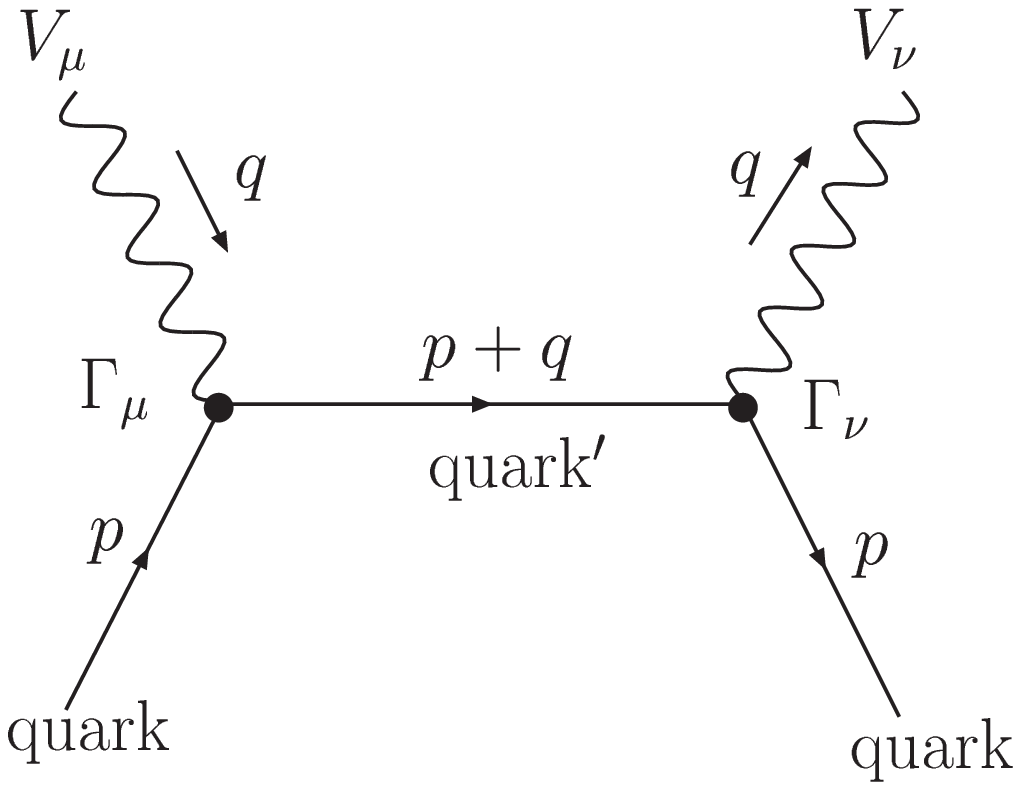}
    \hspace{2.0cm}
    \includegraphics[width=5.0cm]{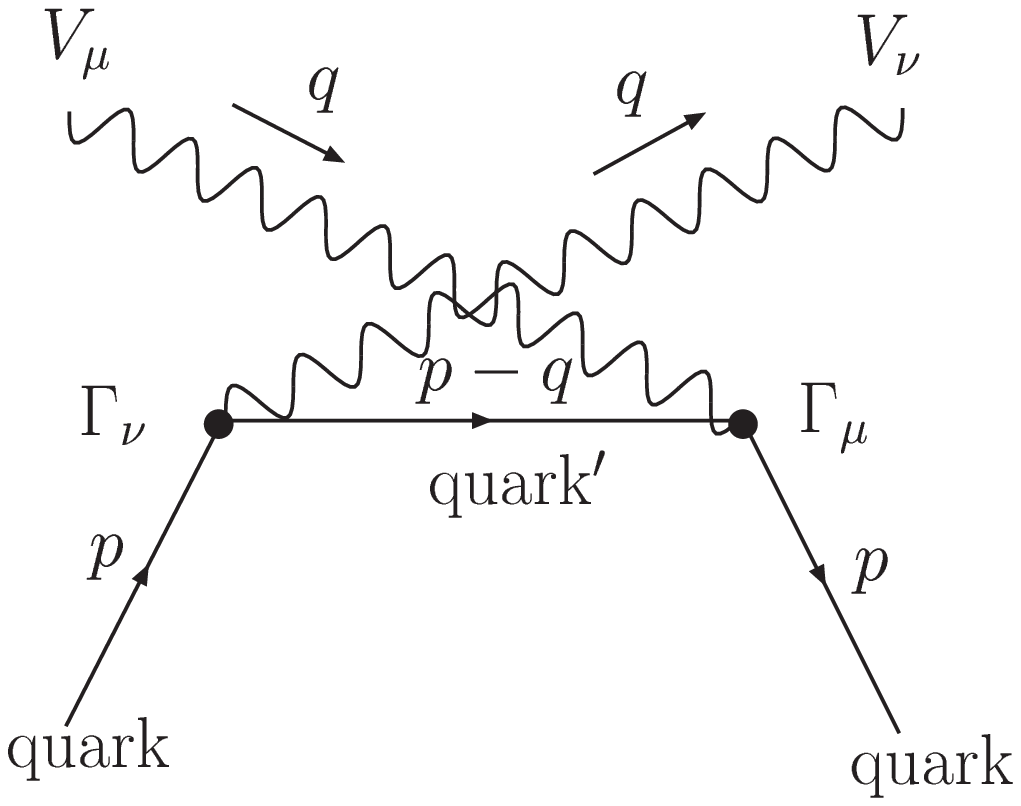}
    \caption[]{\label{fig:crossed-notcrossed} 
      Leading order diagrams contributing to the forward Compton amplitude 
      in deep-inelastic boson$(V)$-quark scattering.}
  \end{center}
\end{figure}
Next, we would like to address the symmetry properties of the partonic forward Compton
amplitude~(\ref{eq:OPEpartons}) $t_{\mu\nu}$ and explain how these translate into 
selection rules for either even or odd Mellin $n$-moments of the different DIS structure functions. 
To that end, 
let us inspect the Feynman diagrams for $t_{\mu\nu}$ at the leading order with initial quarks 
in Fig.~\ref{fig:crossed-notcrossed}.
There the right diagram is simply the crossed diagram of the left one. 
In Fig.~\ref{fig:crossed-notcrossed} we denote the gauge bosons by $V_\mu$ and $V_\nu$. 
For the latter there are various choices as they can either be 
photons $\gamma$ or weak gauge bosons $Z^{0}$ and $W^\pm$. 
The matrix element of the left diagram is proportional to 
\begin{equation}
\label{eq:tmunuvert}
t_{\mu\nu}\propto{}\Gamma_\nu\, \sla{D}(p+q)\, \Gamma_\mu
\, ,
\end{equation}
where $\Gamma_\mu$ and $\Gamma_{\nu}$ denote the vertices of vector boson-fermion coupling, 
while $\sla{D}(p+q)$ is the quark propagator of momentum 
$p+q$, $\sla{D}(p+q)=-1/(\sla{p}+\sla{q})$. 
For the right diagram one has 
\begin{equation}
  \label{eq:tmunuvertc}
t_{\mu\nu}\propto{}\Gamma_\mu\, \sla{D}(p-q)\, \Gamma_\nu
\, .  
\end{equation}
Under the simultaneous transformation $\mu\leftrightarrow\nu$ and $q\rightarrow -q$ 
the matrix element of the crossed diagram is equal to uncrossed one, provided both vertices 
$\Gamma_{\mu}$ and $\Gamma_{\nu}$ have the same structure.

Let us detail this situation for the case of the neutral current DIS first. 
The external bosons $V_\mu$ and $V_\nu$ in Fig.~\ref{fig:crossed-notcrossed} 
being photons $\gamma$ or $Z^{0}$-bosons couple to the vertices  
$\Gamma_{\mu}$ and $\Gamma_{\nu}$. 
The latter are either proportional to $e_{q}\gamma_{\mu}$ and $e_{q}\gamma_{\nu}$
with the fractional quark charge $e_q$ ($\gamma$-boson) or to 
$(v_{f}\gamma_{\mu}+ a_{f}\gamma_{\mu}\gamma_{5})$ and $(v_{f}\gamma_{\nu}+ a_{f}\gamma_{\nu}\gamma_{5})$ 
with the (flavor-depended) vector and axial-vector current coupling constants $v_{f}$ and $a_{f}$ ($Z^{0}$-boson).
In the case of $\gamma-Z^{0}$-interference one has to consider both, 
$\gamma$ and $Z^{0}$, in the initial state in Fig.~\ref{fig:crossed-notcrossed} with a different gauge boson in the final state. 
In the end the effective number of diagrams for the interference contributions
will be doubled. For all neutral current DIS cases the quark flavor, of course,  remains conserved.

At this point, we can relate the action of 
simultaneously transforming $\mu\leftrightarrow\nu$ and $q\rightarrow -q$ 
in all Feynman diagrams contributing to $t_{\mu\nu}$ 
to the parameters of the OPE in Eq.~(\ref{eq:OPEpartons}), 
namely the coefficient functions $C_2$, $C_3$ and $C_L$.
It is clear that the full matrix element for $t_{\mu\nu}$ (l.h.s. of Eq.~(\ref{eq:OPEpartons})) remains unchanged,
since the transformation $\mu\leftrightarrow\nu$ and $q\rightarrow -q$ maps the
crossed and uncrossed diagrams onto each other, even in the case of $\gamma-Z^{0}$-interference 
due to the doubled number of diagrams.
On the r.h.s of the OPE in Eq.~(\ref{eq:OPEpartons}) the tensors $e_{\mu\nu}$
and $d_{\mu\nu}$ remain invariant under $\mu\leftrightarrow\nu$, 
while the antisymmetric tensor $\varepsilon_{\mu\nu\alpha\beta}$ picks up a sign (-1).
The coefficients $C_2$, $C_3$ and $C_L$ as well as the OMEs $A_{\rm parton}$,
being Lorentz scalars, are at most functions of $Q^{2}=-q^{2}$. 
Therefore they are invariant as well.
Finally $\omega$ will be transformed to $-\omega$ (recall its definition $\omega = (2 p\cdot q)/Q^2$).
Thus, in the series expansion in spin $n$ in Eq.~(\ref{eq:OPEpartons})
the coefficient functions $C_2$ and $C_L$ are weighted by a factor $(-1)^{n}$, 
whereas $C_3$ is multiplied by  $(-1)^{n+1}$.
In other words, the sum in Eq.~(\ref{eq:OPEpartons})
runs for $C_2$ and $C_L$ only over even Mellin moments $n$ 
and only over odd for $C_3$. The coefficients for other $n$ have to vanish in Eq.~(\ref{eq:OPEpartons}).
The same choice of $n$ holds for the Mellin moments 
of the structure functions $F_2$ and $F_L$ (even $n$)  Eq.~(\ref{eq:mellindefF2L}) and $F_3$ (odd $n$) Eq.~(\ref{eq:mellindefF3}) of neutral current DIS because of relations  
Eq.~(\ref{eq:F2mellinMR}--\ref{eq:F3mellinMR}) which will be discussed later.

\begin{figure}[ht]
  \begin{center}
    \includegraphics[width=4.0cm]{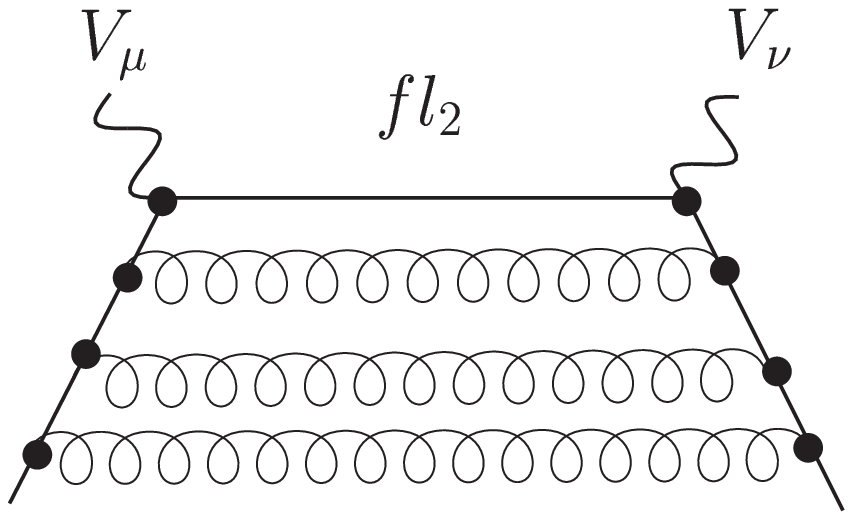}
    \includegraphics[width=4.0cm]{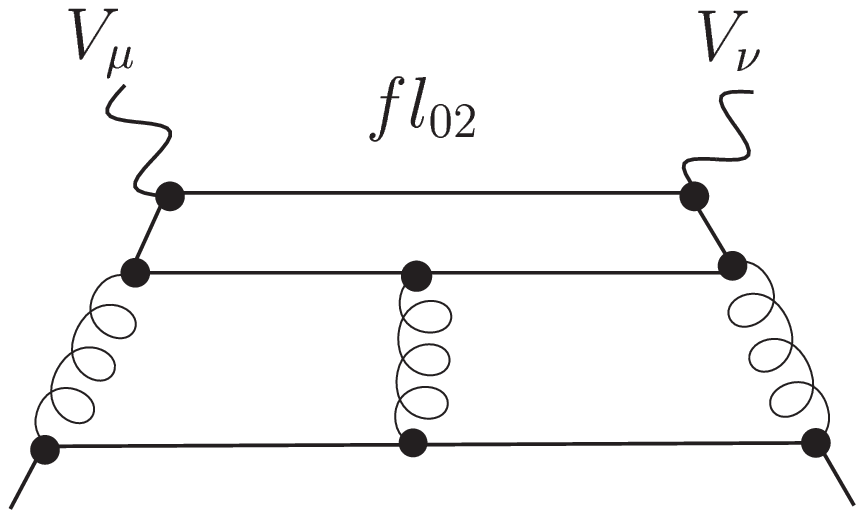}
    \includegraphics[width=4.0cm]{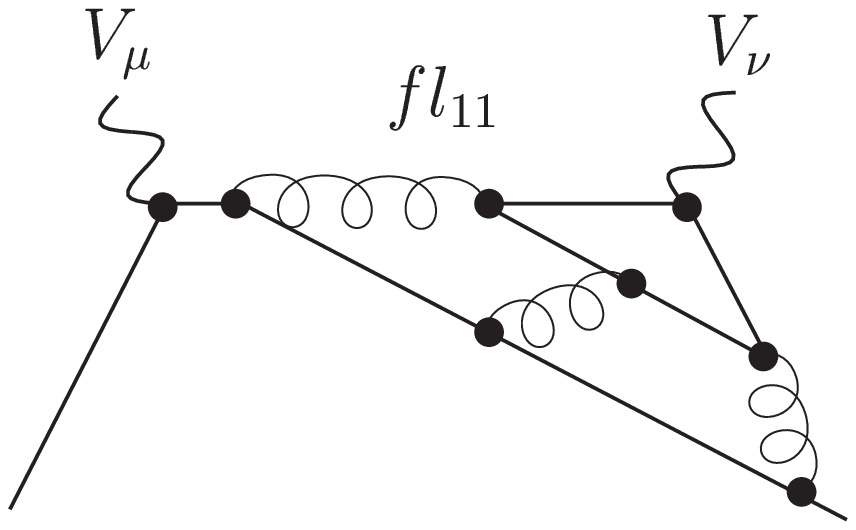}
    \includegraphics[width=4.0cm]{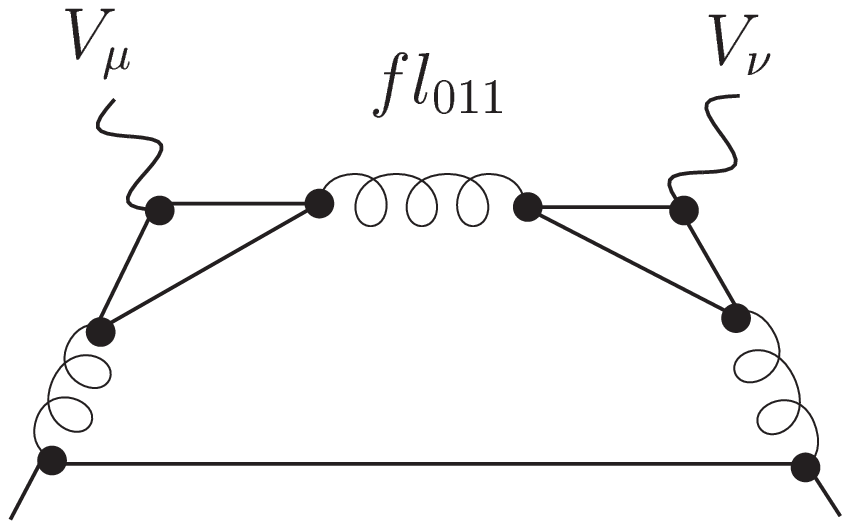}
    \caption[]{\label{fig:various_flavors} Representative three-loop diagrams
      for the various flavor classes in charged current 
      neutrino-proton DIS (see text).}
  \end{center}
\end{figure}
So far our discussion has been based on leading order Feynman diagrams but the
previous arguments carry over to higher orders as well.
Up to three-loop accuracy all Feynman diagrams fall in one of the so-called flavor classes 
$fl_2$, $fl_{02}$, $fl_{11}$ or $fl_{011}$ displayed in Fig.~\ref{fig:various_flavors}.
The class $fl_2$ corresponds to all diagrams with both gauge bosons 
$V_{\mu}$ and $V_{\nu}$ attached to the open fermion line of the initial (final) state quark. 
Class  $fl_{02}$ collects the diagrams with both gauge bosons attached to an
internal closed fermion loop, while $fl_{11}$ contains the
diagrams with one boson attached to the closed loop and the other
to the open line of the external quark. 
Finally the class $fl_{011}$ denotes diagrams with both bosons attached to
different closed quark loops. Depending on the process under consideration some flavor classes vanish identically.
It is easy to see that the neutral current DIS assignments for $C_2$ and $C_L$
(even Mellin moments) and $C_3$ (odd Mellin moments) from above persist and the same holds true for the structure functions. 

\begin{figure}[ht]
  \begin{center}
    \includegraphics[width=5.0cm]{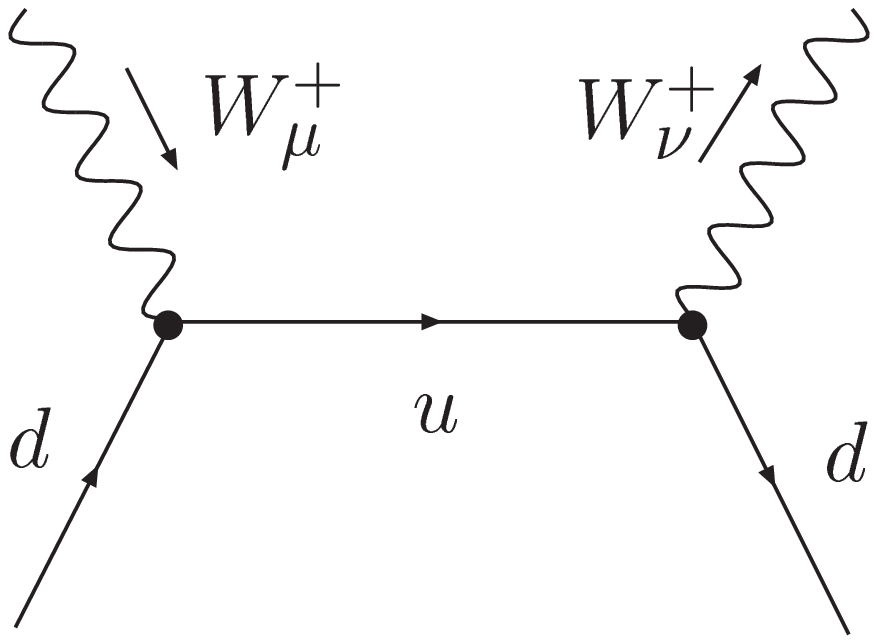}\hspace{2cm}
    \includegraphics[width=5.0cm]{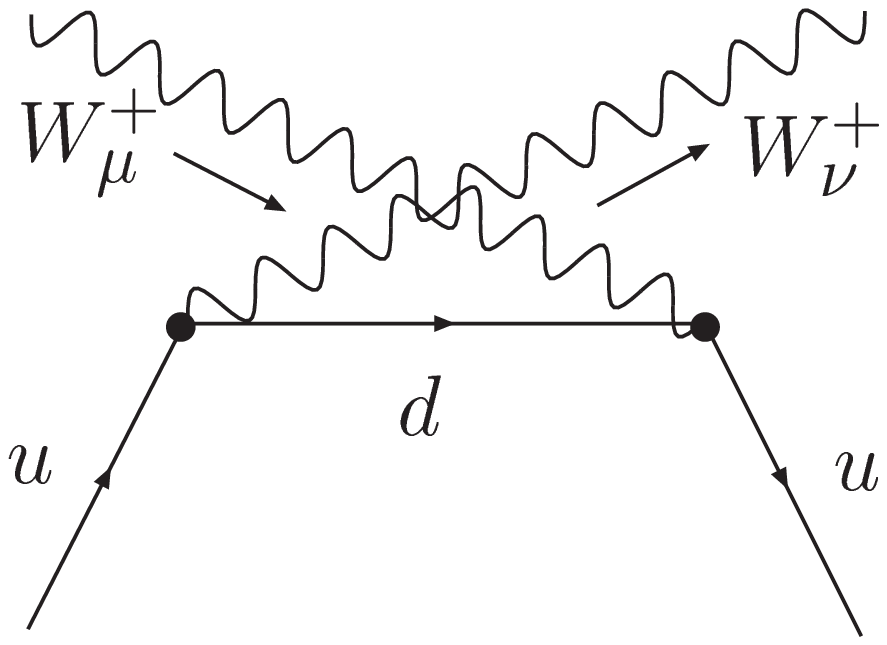}
    \caption[]{\label{fig:u-d-game} Leading order diagrams contributing to the forward
      Compton amplitude of charged current $\nu P$ and $\nu N$ scattering. 
      The right diagram represents the crossed of the left one but with an
      incoming quark of different flavor.}
  \end{center}
\end{figure}
Let us next turn to the case of charged current DIS.
We have the structure functions $F_2$, $F_3$ and $F_L$ for both, an isoscalar
and an isovector target, i.e. ${\nu P \pm \nu N}$, 
which we have to distinguish (see also Eq.~(\ref{eq:FnuPFnuN})).
On the partonic level this implies that we sum the contributions of $u$ and $d$ quarks 
in the singlet case and take their difference in the non-singlet case. 
For simplicity, we restrict ourselves here again to $SU(2)$-isospin symmetry 
with flavors $u$ and $d$ only. 
The generalization to $s$, $c$ and more flavors should be clear. 
In charged current $\nu P$ or $\nu N$ DIS we are considering 
initial and final gauge bosons $V_{\mu}=W^{+}_{\mu}$ and $V_{\nu}=W^{+}_{\nu}$ 
(cf. Fig.~\ref{fig:crossed-notcrossed}), 
the coupling of $d$-quarks to $W^{-}$ being excluded by electroweak theory. 
Say, we take $d$ as incoming and outgoing quark in the left diagram of Fig.~\ref{fig:crossed-notcrossed}.
Then, the scattering of $W^{+}$ with the incoming $d$ quark yields a $u$ quark 
(or $c$ and $t$ if more flavors are considered).
On the other hand, the crossed diagram on the right in Fig.~\ref{fig:crossed-notcrossed} 
simply does not exist for an incoming $d$ quark, 
because it is not allowed by the electroweak Standard Model couplings.
Rather, the incoming quark should be a $u$ quark.
In Fig.~\ref{fig:u-d-game} we display explicitly the appropriate pair of Feynman diagrams 
at leading order.

Thus, for the partonic forward Compton amplitude~(\ref{eq:OPEpartons}) $t_{\mu\nu}$ 
in the combination $t_{\mu\nu}^{u+d}\equiv t_{\mu\nu}^{u}+t_{\mu\nu}^{d}$ 
we effectively sum the contributions of both, crossed and uncrossed, diagrams 
in Fig.~\ref{fig:u-d-game} whereas for $t_{\mu\nu}^{u-d}\equiv t_{\mu\nu}^{u}-t_{\mu\nu}^{d}$ 
we subtract them.
Then, we arrive at the following properties for simultaneous transformations 
$\mu\leftrightarrow\nu$ and $q\rightarrow-q$, 
\begin{eqnarray}
  \label{eq:tmunu-isospin-upd}
t_{\mu\nu}^{u+d}& \rightarrow & t_{\mu\nu}^{u+d}\, ,
\\
  \label{eq:tmunu-isospin-umd}
t_{\mu\nu}^{u-d}&  \rightarrow & (-1)\, t_{\mu\nu}^{u-d}.
\end{eqnarray}

Eq.~(\ref{eq:tmunu-isospin-upd}) implies that the forward Compton amplitude $t_{\mu\nu}^{u+d}$ 
has the same symmetry property as in the case of neutral current DIS. 
For the corresponding coefficient functions and their dependence on the Mellin variable $n$ 
we may repeat exactly the same line of arguments as before leading to the conclusion that 
$C_2$, $C_L$ ($C_3$) are governed by even (odd) $n$ only.
In the other case, Eq.~(\ref{eq:tmunu-isospin-umd}) shows that 
$t_{\mu\nu}^{u-d}$ is antisymmetric under the transformation 
$\mu\leftrightarrow\nu$ simultaneously with $q\rightarrow-q$.
which gives an additional factor $(-1)$ for the l.h.s. of Eq.~(\ref{eq:OPEpartons}).
This alters the Mellin-$n$ dependence of the coefficient functions so that we have 
precisely the opposite assignments, $C_2$, $C_L$ ($C_3$) being entirely odd (even) 
functions of $n$ only.

Before moving on, let us briefly comment on the higher order diagrams for charged current DIS 
as illustrated in Fig.~\ref{fig:various_flavors}. 
For the flavor class $fl_2$ our tree level arguments from above may be literally repeated. 
In the flavor class $fl_{02}$, on the other hand, crossed diagrams with the same external quark 
flavor do contribute. 
However, this does not destroy the symmetry properties of the singlet and non-singlet combinations.
The complete $t_{\mu\nu}^{u+d}$ simply sums the crossed and uncrossed diagrams and, therefore 
is still be symmetric under $\mu \leftrightarrow \nu $ simultaneously with $q\rightarrow-q$, 
thus Eq.~(\ref{eq:tmunu-isospin-upd}) holds for the $u+d$ combination.
In contrast, the contributions from the $u-d$ combination to the flavor class $fl_{02}$ vanish. 
The flavor classes $fl_{11}$ and $fl_{011}$ are excluded in charged current DIS, 
because the flavor changes and, as a consequence, 
the coupling of one single $W^{+}$-boson to a quark loop is not possible.

Finally, it remains to relate the coefficient functions $C_2$, $C_3$ and $C_L$ 
and the OMEs of Eq.~(\ref{eq:OME}) to the Mellin moments of structure functions.
To that end, it is convenient to project Eq.~(\ref{eq:htensor}) and the analogue of Eq.~(\ref{eq:OPEpartons}) for the hadron forward Compton amplitude $T_{\mu\nu}$
onto the respective Lorentz structure using projectors
\begin{eqnarray}
\label{eq:projL}
 P_{L}^{\mu\nu} &\equiv& - \frac{q^2}{(p\mydot q)^2}\, p^\m p^\n \, , 
\\
\label{eq:proj2}
 P_{2}^{\mu\nu} &\equiv& - \left( \frac{3-2\epsilon}{2-2\epsilon}
\hspace{1mm} 
  \frac{ q^2}{(p\mydot q)^2}\, p^\m p^\n
   + \frac{1}{2-2\epsilon} 
\hspace{1mm} 
g^{\m\n} \right) 
\, , 
\\
\label{eq:proj3}
 P_{3}^{\mu\nu}&\equiv& - {\rm{i}}  \frac{1}{(1-2\epsilon)(2-2\epsilon)}
\hspace{1mm} 
\e^{\m\n\a\b}\, \frac{p_\a q_\b}{p\mydot q} 
\, ,
\end{eqnarray}
where all expressions are exact in $D=4-2\varepsilon$ dimensions.

With the help of Eqs.~(\ref{eq:projL})--(\ref{eq:proj3}) one arrives 
at relations between Mellin moments of DIS structure functions~(\ref{eq:mellindefF2L}), (\ref{eq:mellindefF3}) 
and the parameters of OPE. 
On a technical level, this implies a Cauchy integration of the analogue of Eq.~(\ref{eq:OPEpartons}) for $T_{\mu\nu}$ in the complex $\omega$-plane 
 and we recall the necessary details in Appendix~\ref{sec:appB}.
\begin{eqnarray}
  \label{eq:F2mellinMR}
  \int\limits_0^1 dx\, x^{n-2} F_{i}(x,Q^2)
  &=& 
  \sum\limits_{j}
  C_{i,j}\left(n,\frac{Q^2}{\m^2},\a_s\right) 
  A_{ \rm nucl }^{j}\left(n,{\m^2}\right)\, , 
  \quad\quad\quad i=2,L\, , 
  \\
  \label{eq:F3mellinMR}
  \int\limits_0^1 dx\, x^{n-1} F_{3}(x,Q^2)
  &=&
  \sum\limits_{j}
  C_{3,j}\left(n, \frac{Q^2}{\m^2},\a_s\right) 
  A_{\rm nucl}^j\left(n,{\m^2}\right)
  \, .
\end{eqnarray}

To summarize, Eqs.~(\ref{eq:F2mellinMR}) and (\ref{eq:F3mellinMR}) provide the basis 
to obtain Mellin moments of DIS structure functions in our approach 
relying on the OPE and the optical theorem.
Furthermore, from the careful examination of the symmetry properties of the 
forward Compton amplitude $T_{\mu\nu}$ and, related, the underlying Feynman diagrams, 
we have deduced corresponding rules for the Mellin variable $n$.
In the case of neutral current and charged current ${\nu P+\nu N}$ DIS 
the structure functions $F_2$ and $F_L$ are only functions of even Mellin-$n$ 
and only functions of odd $n$ for $F_3$.
For the case of interest in this paper, charged current ${\nu P-\nu N}$ DIS, 
we encounter only odd functions in $n$ for $F_2$ and $F_L$ and, 
only even functions in $n$ for $F_3$, respectively.

%
% ---------------------------------------------------------------------
%
\setcounter{equation}{0}
\section{Renormalization}
\label{sec:renormalization}
%
% ---------------------------------------------------------------------
%

In this Section we briefly recall the necessary steps in renormalizing the
operators in the OPE and, following the discussion above,  
we restrict ourselves here entirely to the non-singlet case.
Starting, say, from the partonic expression~(\ref{eq:OPEpartons}), 
i.e. partonic matrix elements of $t_{\mu\nu}$, 
we express the renormalized OMEs $A^{j}_{\rm parton}$ (see Eq.~(\ref{eq:OMEpartons})) 
in terms of matrix elements of bare composite operators,
\begin{equation}
\label{eq:Orenns}
  O^{\alpha,{\rm  \,ren }} \: = \: Z_{\rm ns}\,O^{\alpha, \,{\rm bare}} 
  \, .
\end{equation}
Here and later we suppress other indices and the explicit dependence on $n$ for the operators~(\ref{eq:defoperatorns}) 
The scale dependence of the operator $O^{\alpha}$ is governed by the anomalous dimension $\gamma_{\rm ns}$, 
\begin{equation}
\label{eq:gamma_ns}
  \frac{d}{d \ln \mu^2 }\, O^{\alpha, {\rm\, ren} } \:\equiv \: - \,\gamma_{\rm ns}\, O^{\alpha , {\rm\, ren}}  
  \, ,
\end{equation}
and is connected to the renormalization constant $Z_{\rm ns}$ 
in Eq.~(\ref{eq:Orenns}) by
\begin{equation}
  \label{eq:gamZns}
  \gamma_{\rm ns} \: = \: -\,\left( \frac{d }{d\ln\mu^2 }\, Z_{\rm ns} \right) Z^{-1}_{\rm ns} 
  \, .
\end{equation}

In order to arrive at explicit expressions for $Z_{\rm ns}$ or Eq.~(\ref{eq:gamZns}), 
one has to make use of a regularization procedure and a renormalization scheme.
We choose dimensional regularization~\cite{'tHooft:1972fi,Bollini:1972ui,Ashmore:1972uj,Cicuta:1972jf} 
in $D = 4 - 2\ep$ dimensions and the modified minimal subtraction 
\cite{'tHooft:1973mm,Bardeen:1978yd} scheme, $\overline{\rm{MS}}$.
The running coupling evolves according to
\begin{equation}
\label{eq:arun}
  \frac{d}{d \ln \mu^2}\: \frac{\as}{4\pi} \:\: \equiv \:\: 
  \frac{d\,\ar}{d \ln \mu^2} \:\: = \:\: - \ep\, \ar 
  - \beta_0\, a_{\rm s}^2 - \beta_1\, a_{\rm s}^3 
  - \beta_2\, a_{\rm s}^4 - \ldots \:\: ,
\end{equation}
and we have introduced the common short hand notation $a_{s}\equiv \alpha_{s}/(4 \pi)$. 
The usual four-dimensional expansion coefficients $\beta_{\rm n}$ of the beta
function in QCD read $\,\beta_0 = 11 - 2/3\,\nf\,$ etc, 
with $\nf$ representing the number of active quark flavors.
The bare and the renormalized coupling, 
$\a^{\rm{bare}}_s$ and $\a_s$ are related by 
\begin{eqnarray}
\label{eq:alpha-s-renorm}
\a^{\rm{bare}}_s &=& 
Z_{\a_s}\, \a_s^{\rm{}}\, ,
\end{eqnarray}
where we have put the factor $S_\ep = \exp( \ep\{\ln(4\p) - \g_{\rm{E}}\}) = 1$ 
in the $\overline{\rm{MS}}$-scheme and the renormalization constant $Z_{\a_s}$ reads 
\begin{eqnarray}
\label{eq:Z-alpha-s}
Z_{\a_s} &=& 
1 - \frac{\b_0}{\ep} {a_s}\, 
+ \left( \frac{\b_0^2}{\ep^2} - \frac{\b_1}{2\ep} \right) 
{a_s}^2 + 
\dots\,\, .
\end{eqnarray}

In this framework, the renormalization factor $Z_{\rm ns}$ in Eq.~(\ref{eq:Orenns}) 
is a series of poles in $1/\ep$,
expressed in terms of $\beta_{\rm n}$ and the coefficients 
$\gamma^{\,(l)}$ of the anomalous dimensions from an expansion in $\ar$, 
\begin{equation}
\label{eq:gam-exp}
  \gamma(n) \: = \: \sum_{l=0}^{\infty}\,
  a_{\rm s}^{\,l+1}\, \gamma^{\,(l)}(n) 
  \, .
\end{equation}
Up to the third order in the coupling constant 
the expansion of $Z_{\rm ns}$ reads
\begin{eqnarray}
\label{eq:Zns3}
  Z_{\rm ns} & = & 
    1 \: + \: \:\ar\, \frac{1}{\ep}\,\gamma_{\,\rm ns}^{\,(0)} 
    \: + \: a_{\rm s}^2 \,\left[\, \frac{1}{2\ep^2}\, 
    \left\{ \left(\gamma_{\,\rm ns}^{\,(0)} - \beta_0 \right) 
    \gamma_{\,\rm ns}^{\,(0)} \right\}
    + \frac{1}{2\ep}\, \gamma_{\,\rm ns}^{\,(1)} \right] 
  \nonumber \\[1mm] & & \mbox{} + \:  
  a_{\rm s}^3 \,\left[\, \frac{1}{6\ep^3}\, 
    \left\{ \left( \gamma_{\,\rm ns}^{\,(0)} - 2 \beta_0 \right)
    \left( \gamma_{\,\rm ns}^{\,(0)} - \beta_0 \right) 
    \gamma_{\,\rm ns}^{\,(0)} \right\} \right.
  \nonumber \\[1mm] & & \left. \mbox{} \quad\quad \! + \:  
  \frac{1}{6\ep^2}\, \left\{ 3\, \gamma_{\,\rm ns}^{\,(0)}  
    \gamma_{\,\rm ns}^{\,(1)} - 2 \beta_0\, \gamma_{\,\rm ns}^{\,(1)}
    - 2 \beta_1\, \gamma_{\,\rm ns}^{\,(0)} \right\} \: + \: 
    \frac{1}{3\ep}\, \gamma_{\,\rm ns}^{\,(2)} \right] \:\: .
\end{eqnarray}
The anomalous dimensions $\gamma^{\,(l)}$ can thus be read off from the
$\ep^{-1}$ terms of the renormalization factors at order $a_{\rm s}
^{\,l+1}$, while the higher poles in $1/\ep$ can serve as checks for 
the calculation. 
The coefficient functions in Eqs.~(\ref{eq:OPE}), (\ref{eq:OPEpartons}), 
on the other hand, have an expansion in positive powers of $\ep$, 
\begin{equation}
\label{eq:cf-exp}
  C_{i,{\rm ns}} \: = \: \sum_{l=0}^{\infty} \, a_{\rm s}^{\, l} \left( c_{i,{\rm ns}}^{\,(l)} 
    + \ep a_{i,{\rm ns}}^{\,(l)} + \ep^2 b_{i,{\rm ns}}^{\,(l)} + \ldots \right)
  \, ,
\end{equation}
where $i =2, 3, L$ and we have again suppressed the dependence on $n$ (and $Q^2/\mu^2$). Here $C_{i,{\rm ns}}$ is our generic 
notation for non-singlet contributions obtained for $C_{i,\alpha}$ in Eqs.~(\ref{eq:OPE}), (\ref{eq:OPEpartons})

Due to the presence of $\gamma_5$ in the vertices, 
the axial-vector coupling in dimensional regularization brings up the 
need for additional renormalizations to restore the axial Ward-identities.
This is extensively described in the literature and for the associated renormalizations 
we use the prescription of Ref.~\cite{Larin:1991tj,Larin:1993tq} based on relating 
vector and axial-vector currents. 
The necessary constant $Z_A$ for the axial renormalization 
$Z_5$ and the finite renormalization due to the treatment of $\gamma_5$ 
in the $\overline{\rm{MS}}$-scheme are known to three loops~\cite{Larin:1991tj,Larin:1993tq}.

The actual calculation of the anomalous dimension~(\ref{eq:gamma_ns}) 
and the coefficient functions $C_{i,{\rm ns}}$ in perturbative QCD proceeds as follows. 
Using the Lorentz projectors~(\ref{eq:projL})--(\ref{eq:proj3}) we obtain 
from the forward partonic Compton amplitude Eq.~(\ref{eq:OPEpartons}) 
the partonic invariants 
\begin{equation}
  \label{eq:def-partoninv}
  t_{i,{\rm ns} }=P^{\mu\nu}_{i} t_{\mu\nu}
  \, , 
  \quad\quad\quad
  i=2,3,L
  \, , 
\end{equation}
see also Eqs.~(\ref{eq:invarTmunu2L}), (\ref{eq:invarTmunu3}). 
These invariants can be written in terms of the bare operator matrix elements as
\begin{eqnarray}
\lefteqn{
\label{eq:TmunuPartonRenNS}
t_{ i,{\rm ns} } (x,Q^2,\a_s,\ep) \,=} \\
& &
2 \sum_{n}
\left( \frac{1}{x}\right)^n 
C^{{\rm{}}}_{i,{\rm ns}}\left(n,\frac{Q^2}{\m^2},\a_s,\ep\right) 
Z_{\rm{ns}}\left(\a_s,\frac{1}{\ep}\right)
A^{ {\rm ns} ,{\rm bare } }_{{\rm q}}\left(n,\a_s,\frac{p^2}{\m^2},\ep\right) 
+ O(p^2)
\, , \nonumber
\end{eqnarray}
where $i=2,3,L$ and 
the l.h.s. of Eq.~(\ref{eq:TmunuPartonRenNS}) 
is renormalized by substituting the bare coupling constant
in terms of the renormalized, see Eq.~(\ref{eq:alpha-s-renorm}).
The wave function renormalization for the external quark lines is an overall factor 
on both sides of the Eq.~(\ref{eq:TmunuPartonRenNS}) and drops out.
The terms $O(p^2)$ on the r.h.s. of Eq.~(\ref{eq:TmunuPartonRenNS}) 
indicate higher twist contributions, which we neglect.

Starting with the partonic invariant $t_{i,{\rm{ns}}}$ 
from Eq.~(\ref{eq:TmunuPartonRenNS}), the renormalization constants $Z_{\rm ns}$ 
and the coefficient functions  $C_{i,{\rm ns}}$ 
are calculated using the method of projection developed in Ref.~\cite{Gorishnii:1983su}, 
which consists of applying the following projection operator, 
\begin{eqnarray}
\label{eq:projectionoperator}
 {\cal P}_n[f(p,q)] \equiv
  \left. \Biggl[ \frac{q^{ \{\m_1}\cdots q^{\m_n \}}}{2 n !}
  \frac{\partial ^n}{\partial
p^{\m_1} \cdots  \partial p^{\m_n}} f(p,q) \Biggr] \right|_{p=0} \, ,
\end{eqnarray}
to both sides of Eq.~(\ref{eq:TmunuPartonRenNS}). 
Here $q^{ \{\m_1}\cdots q^{\m_n \}}$ is symmetrical and traceless, 
i.e. the harmonic part of the tensor $q^{\m_1}\cdots q^{\m_n }$.

On the r.h.s. of Eq.~(\ref{eq:TmunuPartonRenNS}), it is obvious, 
that the $n$-th order differentiation in the projection operator ${\cal P}_n$ 
singles out precisely the $n$-th moment, i.e. the coefficient of $1/x^{n}$. 
All other powers of $1/x$ vanish either by differentiation or after nullifying 
the momentum $p$. 
The operator ${\cal P}_n$ does not act on the 
renormalization constant $Z_{\rm ns}$ and the coefficient functions 
on the r.h.s. of Eq.~(\ref{eq:TmunuPartonRenNS}) 
as they are only functions of $n$, $\a_s$ and $\ep$.
However, ${\cal P}_n$ does act on the partonic bare OMEs $A^{j,{\rm bare}}_{{\rm{parton}}}$, 
where the nullification of $p$ effectively eliminates all 
but the tree level diagrams $A^{j,{\rm{tree}}}_{{\rm parton}}$, 
because any diagram with loops becomes a massless tadpole 
and is put to zero in dimensional regularization.
Finally, the $O(p^2)$ terms in Eq.~(\ref{eq:TmunuPartonRenNS}), 
which denote higher twist contributions,  
become proportional to the metric tensor after differentiation. 
They are removed by the harmonic tensor $q^{ \{\m_1}\cdots q^{\m_n \}}$.
On the l.h.s. of Eq.~(\ref{eq:TmunuPartonRenNS}),
${\cal P}_{n}$ is applied to the integrands of all Feynman diagrams 
contributing to the invariants $t_{i,{\rm{parton}}}$. 
The momentum $p$ is nullified before taking the limit $\ep \rightarrow 0$, 
so that all infrared divergences as $p\rightarrow 0$ are dimensionally 
regularized for individual diagrams. 
This reduces the 4-point diagrams that contribute to $t_{\mu\nu}$ 
to self-energy type diagrams (2-point-functions) accessible to reduction algorithms such 
as {\sc Mincer}~\cite{Larin:1991fz} (see the Section~\ref{sec:calculation}).

To summarize, we find after application of the projection operator
${\cal P}_n$ to Eq.~(\ref{eq:TmunuPartonRenNS}) 
\begin{eqnarray}
\label{eq:TmunuPartonMomNS}
t_{i,{\rm{ns}}} \left(n,\frac{Q^2}{\m^2},\a_s,\ep\right)
&\equiv&
\cp_n\, t_{i,\rm ns}(x,Q^2,\a_s,\ep)
\\
&=&
C_{i,{\rm ns}}\left(n,\frac{Q^2}{\m^2},\a_s,\ep\right) 
Z_{\rm{ns}}\left(\a_s,\frac{1}{\ep}\right)
A^{{\rm ns},{\rm tree}}_{\rm q}(n,\ep)  
\, , 
\nonumber
\end{eqnarray}
where $i = 2,3,L$.
This is our starting point for an iterative determination of the anomalous dimensions and 
coefficient functions via the OPE, since the $C_{i,{\rm ns}}$ ($Z_{\rm{ns}}$) are
expanded in positive (negative) powers of $\ep$ while the OME $A_{{\rm q}}^{\rm ns, tree}$ 
does factorize after application of the projector $\cp_n$. 
In a series expansion in terms of the renormalized coupling $\ar$ 
at the scale $\mu^2 = Q^2$ we can write 
\begin{eqnarray}
\label{eq:partinv-exp}
t_{i,{\rm ns}}(n) &=& \left( 
t^{(0)}_{i,{\rm ns}}(n) +a_{s}\, t^{(1)}_{i,{\rm ns}}(n) 
+
a_{s}^{2}\, t^{(2)}_{i,{\rm ns}}(n) 
+ 
a_{s}^{3}\, t^{(3)}_{i,{\rm ns}}(n) 
+ \dots 
\right) A^{{{\rm ns}},{\rm{tree}}}_{{\rm q}}(n)
\, ,
\end{eqnarray}
with $i = 2,3,L$ and recall that we use $a_{s}\equiv \alpha_{s}/(4 \pi)$. 
Then we normalize leading order contribution as follows, 
\begin{eqnarray}
\label{eq:F-0}
t^{(0)}_{i,{\rm ns}}(n) = 1
\, ,
\quad
i = 2,3
\, ,
\quad\quad
\mbox{and}
\quad\quad 
t^{(0)}_{L,{\rm ns}}(n) = 0
\, ,
\end{eqnarray}
where the OME $A_{{\rm q}}^{\rm ns, tree}$ (being a constant) has been absorbed into 
the normalization of Eq.~(\ref{eq:F-0}). 
With the normalization~(\ref{eq:F-0}) one has 
\begin{eqnarray}
\label{eq:F2L3-NS-0}
c^{(0)}_{i,{\rm ns}}(n)=1
\, ,
\quad
i = 2,3
\, ,
\quad\quad
\mbox{and}
\quad\quad 
c^{(0)}_{L,{\rm ns}}(n)=0
\, .
\end{eqnarray}
At first order in $\a_s$, expanding up to order $\ep^{2}$ 
and suppressing the $n$-dependence from now on for brevity, we find 
\begin{eqnarray}
  \label{eq:F2-NS-1}
  t^{(1)}_{i,{\rm ns}} &=& 
    \frac{1}{\ep} \g^{(0)}_{\rm ns} 
  + c^{(1)}_{i,{\rm ns}}
  + \ep a^{(1)}_{i,{\rm ns}} + \ep^{2} b^{(1)}_{i,{\rm ns}} +{\cal O}(\ep^{3})
\, ,
\quad
i = 2,3
  \, , \\[1ex]
\label{eq:FL-NS-1}
t^{(1)}_{L,{\rm ns}} &=& 
     c^{(1)}_{L,{\rm ns}}               
   + \ep a^{(1)}_{L,{\rm ns}} + \ep^{2} b^{(1)}_{L,{\rm ns}} +{\cal O}(\ep^{3})
   \, . 
\end{eqnarray}
Performing the expansion at $\alpha_{s}^{2}$ up to order $\ep$ we arrive at the equations:
\begin{eqnarray}
\label{eq:F2-NS-2}
t^{(2)}_{i,{\rm ns}} &=& 
      \frac{1}{2 \ep^2} 
    \left\{ \left(  \g^{(0)}_{\rm ns } - \b_0  \right)  \g^{(0)}_{\rm ns }
    \right\} 
    + \frac{1}{2 \ep} \left\{ 
      \g^{(1)}_{\rm ns }
      + 2  c^{(1)}_{i,{\rm ns}}\, \g^{(0)}_{\rm ns}
    \right\} 
\\
& &
    + c^{(2)}_{i,{\rm ns}} +  a^{(1)}_{i,{\rm ns}}\,  \g^{(0)}_{\rm ns} 
    + \ep \left\{  a^{(2)}_{i,{\rm ns}}
      +  b^{(1)}_{i,{\rm ns}}\, \g^{(0)}_{\rm ns }  
    \right\}  +{\cal O}(\ep^{2})
\, ,
\quad
i = 2,3
    \, , 
\nonumber
\\[1ex]
\label{eq:FL-NS-2}
t^{(2)}_{L,{\rm ns}} &=& 
      \frac{1}{ \ep} \left\{
      c^{(1)}_{L,{\rm ns}}\, \g^{(0)}_{\rm ns} \right\} 
    + c^{(2)}_{L,{\rm ns}} +  a^{(1)}_{L,{\rm ns}}\,  \g^{(0)}_{\rm ns} 
    + \ep \left\{  a^{(2)}_{L,{\rm ns}}
    + b^{(1)}_{L,{\rm ns}}\, \g^{(0)}_{\rm ns } \right\} 
    + {\cal O}(\ep^{2})
\, .
\nonumber
\end{eqnarray}
Finally for the third order $\alpha_{s}^{3}$  we obtain
\begin{eqnarray}
\label{eq:F2-NS-3}
t^{(3)}_{i,{\rm ns}} &=&  
      \frac{1}{6 \ep^3} 
    \left\{ \left( \g^{(0)}_{\rm ns} - 2 \b_{0} \right)\left( \g^{(0)}_{\rm ns}-\b_{0}  \right)
      \g^{(0)}_{\rm ns} \right\} 
\\
& &
      \frac{1}{6 \ep^2} 
    \left\{ 3  \g^{(0)}_{\rm ns}\,  \g^{(1)}_{\rm ns}-  2 \b_{0}\,
      \g^{(1)}_{\rm ns} - 2 \b_{1}\, \g^{(0)}_{\rm ns} + 3c^{(1)}_{i,{\rm ns}}\left( \g^{(0)}_{\rm ns}-\b_{0}\right)  \g^{(0)}_{\rm ns}
    \right\} 
\nonumber \\
& &
    + \frac{1}{6  \ep} \left\{2 \g^{(2)}_{\rm ns} +3 c^{(1)}_{i,{\rm ns}}\,
      \g^{(1)}_{\rm ns} + 6 c^{(2)}_{i,{\rm ns}}\,  \g^{(0)}_{\rm ns}
    + 3 a^{(1)}_{i,{\rm ns}} \left( \g^{(0)}_{\rm ns}-\b_{0}\right)\g^{(0)}_{\rm ns}                                
    \right\}
\nonumber \\
& &
    + \frac{1}{2}\left\{ 2 c^{(3)}_{i,{\rm ns}} + a^{(1)}_{i,{\rm ns}}\, \g^{(1)}_{\rm ns} 
      + 2  a^{(2)}_{i,{\rm ns}}\, \g^{(0)}_{\rm ns} + b^{(1)}_{i,{\rm ns}}\left(
        \g^{(0)}_{\rm ns} - \b_{0}\right)\, \g^{(0)}_{\rm ns} 
    \right\}
    + {\cal O}(\ep)
\, ,
\quad
i = 2,3
\, , 
\nonumber 
\\[1ex]
\label{eq:FL-NS-3}
t^{(3)}_{L,{\rm ns}} &=&  
      \frac{1}{2 \ep^2} \left\{ c^{(1)}_{L,{\rm ns}}\left( \g^{(0)}_{\rm ns}-\b_{0}\right)  \g^{(0)}_{\rm ns}
    \right\}
\\
& &
    + \frac{1}{2  \ep} \left\{ c^{(1)}_{L,{\rm ns}}\,  \g^{(1)}_{\rm ns} +2  c^{(2)}_{L,{\rm ns}}\,  \g^{(0)}_{\rm ns}
    +  a^{(1)}_{L,{\rm ns}} \left( \g^{(0)}_{\rm ns}-\b_{0}\right)\g^{(0)}_{\rm ns} 
  \right\}\nonumber \\
& &
   + \frac{1}{2}\left\{ 2 c^{(3)}_{L,{\rm ns}} + a^{(1)}_{L,{\rm ns}}\, \g^{(1)}_{\rm ns} 
     + 2  a^{(2)}_{L,{\rm ns}}\, \g^{(0)}_{\rm ns}+ b^{(1)}_{L,{\rm ns}}\left(
       \g^{(0)}_{\rm ns} - \b_{0}\right)\, \g^{(0)}_{\rm ns} \right\}
   + {\cal O}(\ep)
\, .
\nonumber 
\end{eqnarray}

Eqs.~(\ref{eq:F2-NS-1})--(\ref{eq:FL-NS-3}) hold for both, even and odd Mellin
moments alike and we did not distinguish these in our notation.
However, from the discussion of the preceding Sections it is clear that the 
respective anomalous dimensions $\gamma^{(l)}$ and coefficient functions $c^{(l)}_{i,{\rm ns}}$
describe different physical processes.
In fact, it is well-known that starting from 
$\gamma^{(1)}$ and $c^{(2)}_{i,{\rm ns}}$ (and $a^{(2)}_{i,{\rm ns}}$, $b^{(2)}_{i,{\rm ns}}$, etc.), 
they differ.
The new results of the present paper from Eqs.~(\ref{eq:F2-NS-3}), (\ref{eq:FL-NS-3}) 
at third order in $\alpha_{s}$ consist of odd Mellin moments for $c^{(3)}_{2,{\rm ns}}$ 
and $c^{(3)}_{L,{\rm ns}}$ and even moments for $c^{(3)}_{3,{\rm ns}}$. 
Below in Sec.~\ref{sec:calculation} we present numerical results for them,
while complete expressions through rational numbers 
are deferred to Appendix~\ref{sec:appA}.

%
% ---------------------------------------------------------------------
%
\setcounter{equation}{0}
\section{Calculation}
\label{sec:calculation}
%
% ---------------------------------------------------------------------
%

In the previous Sections, we have laid the foundations to our calculation of
Mellin moments of the DIS charged current structure functions 
$F_2^{\nu P -\bar \nu P}$, $F_3^{\nu P -\bar \nu P}$ and $F_L^{\nu P -\bar \nu P}$ 
together with their respective coefficient functions and anomalous dimensions.
To that end, following Refs.~\cite{Moch:2004pa,Vogt:2004mw,Moch:2004xu,Vermaseren:2005qc}, 
we have calculated the Lorentz invariants of the parton Compton
amplitude $t^{(l)}_{i,{\rm ns}}$, $l=0,1,2,3$, $i=2,3,L$, 
as given in the l.h.s. of Eqs.~(\ref{eq:F2-NS-1})--(\ref{eq:FL-NS-3}) from first principles. 
All contributing Feynman diagram were generated and  
then projected by one of the Lorentz projection~(\ref{eq:projL})--(\ref{eq:proj3}). 
Subsequently, the application of Eq.~(\ref{eq:projectionoperator}) for the harmonic projection ${\cal P}_n$ 
extracts all contributions to the given Mellin moment, which are finally
solved in terms of rational numbers, values of the Riemann zeta functions and
$SU(N_c)$ color coefficients $C_A$, $C_F$ and $n_f$.
Due to the large number of diagrams involved in the calculations up to order $\alpha_{s}^{3}$ 
sufficient automatization is necessary.
Therefore the calculations are organized in detail as follows:
\begin{table}
\begin{center}
\begin{tabular}{ccccccc}    \hline\hline
 & & & & & & \\[-3mm]
Lorentz     &Structure  & tree & one-loop & two-loop &three-loop & sum\\
invariant  &function   &$ {\cal O}(\alpha_{s}^{0})$ & ${\cal O}(\alpha_{s}^{1})$ &  ${\cal O}(\alpha_{s}^{2})$ & ${\cal O}(\alpha_{s}^{3})$ &   \\
 & & & & & & \\[-3mm]
 \hline
 & & & & & & \\[-4mm]
$t_{2,{\rm ns}}$ & $F_{2}^{ \nu P - \bar \nu P}$ & 1 & 4 & 55 & 1016 & 1076  \\
 & & & & & & \\[-4mm]
\hline
 & & & & & & \\[-4mm]
$t_{L,{\rm ns}}$ & $F_{L}^{ \nu P - \bar \nu P}$ & 1 & 4 & 55 & 1016 & 1076 \\
 & & & & & & \\[-4mm]
\hline
 & & & & & & \\[-4mm]
$t_{3,{\rm ns}}$ & $F_{3}^{ \nu P - \bar \nu P}$ & 1 & 4 & 63 & 1246 & 1314   \\
 & & & & & & \\[-4mm]
\hline\hline
 & & & & & & \\[-3mm]
in total & & & & & & 3466\\
 & & & & & & \\[-3mm]
 \hline  \hline
\end{tabular}
\caption{The number of diagrams involved in the calculation of the 
$\nu P - \bar \nu P$ charged current DIS structure functions $F_2$, $F_L$ 
and $F_3$ at tree level, one-loop, two-loop and three-loop, respectively.}
\label{t:tab1}
\end{center}
\end{table}

\begin{itemize}
\item
  All Feynman diagrams are generated automatically with the program {\sc Qgraf}~\cite{Nogueira:1991ex}. 
  This program generates all possible Feynman diagrams (and topologies) 
  for a given process in some special format.
  The program works very effectively, producing a database with thousands of
  diagrams within seconds only.
  For charged current DIS we have obtained from {\sc Qgraf} 2, 10, 153 and 3468
  diagrams for the tree, one-loop, two-loop and three-loop contributions, 
  respectively.
  
\item For all further calculations we have relied on the latest version of 
  the symbolic manipulation program {\sc Form}~\cite{Vermaseren:2002rp,Vermaseren:2006ag}.
  For the further treatment of {\sc Qgraf} output, such as analysis of the 
  topologies, the explicit implementation of Feynman rules etc. 
  we have adapted a dedicated {\sc Form} procedure {\it conv.prc} from 
  previous work, e.g. Ref.~\cite{Vermaseren:2005qc}. 
  Most importantly, this procedure tries to exploit as many symmetry properties 
  of the original Feynman diagrams in order to reduce their total number.
  The upshot of these efforts are presented in Table~\ref{t:tab1} order by order 
  for structure function corresponding to the different Lorentz projections.
  As one can see, the number of diagrams obtained for $F_3^{\nu P -\bar \nu P}$ is always larger 
  than for $F_2^{\nu P -\bar \nu P}$ or $F_L^{\nu P -\bar \nu P}$. 
  The reason is that in the former case we can not apply certain symmetry
  transformations due to the presence of $\gamma_{5}$ in the vertices.
  The database for $F_2^{\nu P -\bar \nu P}$ and $F_L^{\nu P -\bar \nu P}$ 
  produced by us does almost coincide with the one used in Ref.~\cite{Retey:2000nq}, 
  except for small modifications. The database for $F_3^{\nu P -\bar \nu P}$ is completely new.

\item For the calculation of the color factors for each Feynman diagram 
  we have used the {\sc Form} package {\it color.h}~\cite{vanRitbergen:1998pn}.

\item The actual calculation of the Mellin moments of the Feynman integrals
  has made use of {\sc Mincer}. The detailed description of this program can be found in
  Ref.~\cite{Larin:1991fz} for the {\sc Form} package {\it mincer.h}.
  For organization of the work of (a slightly modified version of) {\sc Mincer}
  with the input databases we have used a dedicated database program {\sc Minos}~\cite{Larin:1994vu,Larin:1997wd}.

\item Finally, on top of {\sc Mincer} and {\sc Minos} some shell scripts
  managed the automatic runs of both programs for different parts of the calculation. 
  This facilitates the bookkeeping of different input parameters for {\sc Minos} and {\sc Mincer}
  due to different Lorentz projections, orders of $\alpha_{s}$ etc in
  distributed running. 
  Moreover, the shell scripts also organized the final summations over 
  the flavor classes as well as the output of all final results.
\end{itemize}

Next, let us discuss the various checks we performed on the results of our calculations.
First of all, we have tested our set-up by a recalculation of some known even
Mellin moments for $F_2$, $F_L$ and odd moments for $F_3$ to find agreement 
with the published results of 
Refs.~\cite{Larin:1994vu,Larin:1997wd,Retey:2000nq,Vermaseren:2005qc}.
Then, most importantly, we have checked gauge invariance, 
i.e. we calculated all our diagrams for all results for all Mellin moments presented 
in this article with a gauge parameter $\xi$ for the gluon propagator,
\begin{equation}
\label{eq:gluonprop}
{\rm i} \frac{-g^{\mu\nu}+(1-\xi) q^{\mu } q^{\nu}}{q^2}.
\end{equation}
We kept all powers of $\xi$ (up to $\xi^{4}$ in three loops for this calculation) through the entire calculation. 
Since parton structure functions are physical quantities any dependence on the
gauge parameter $\xi$ must disappear in the final result. 
This was indeed the case after summing all diagrams in a given flavor class.
Furthermore, we have compared the anomalous dimensions 
$\gamma_{\rm ns}$~Eq.~(\ref{eq:gamma_ns}) as calculated by us from Eqs.~(\ref{eq:F2-NS-1})--(\ref{eq:FL-NS-3}) 
up to three loops with the results available in the
literature~\cite{Moch:2004pa} and found complete agreement for both, even and odd Mellin moments.
In addition, the coefficient functions for the structure functions
$F_{i}^{ \nu P - \bar \nu P},\,\, i=2,3,L$ at two loops are known from earlier
work. Our two-loop results as obtained from Eq.~(\ref{eq:F2-NS-1})--(\ref{eq:FL-NS-2})
coincide with Refs.~\cite{vanNeerven:1991nn,Zijlstra:1991qc,Zijlstra:1992kj,Zijlstra:1992qd,Moch:1999eb}.

Finally, let us mention a few words on the hardware requirements. 
All calculations are CPU-time and disk space consuming, 
especially for the higher Mellin moments (higher $n$ values).
They were typically performed on an 64-bit AMD Opteron 2.2 GHz Linux machine with 4 GByte of memory. 
For example, the calculation of $t_{3,{\rm ns}}$ for $n=10$ took 56
days with the gauge parameter included, while the calculation of both, 
$t_{2,{\rm ns}}$ and $t_{L,{\rm ns}}$ for $n=9$  on the
same machine needed 33 days.
For comparison, the calculation of lowest Mellin moment $n=1$ for both projections
$t_{2,{\rm ns}}$ and $t_{L,{\rm ns}}$ consumes less than a hour, whereas for 
$n=2$ for  $t_{3,{\rm ns}}$ one needs approximately a couple of
hours, always  with the full gauge parameter dependence.
At intermediate stages the calculations required also a large amount of disk space. 
Although the programs calculate diagrams one by one at the time the size of intermediate algebraic expressions 
for some diagrams can grow up to 20 GByte of a disk space (for instance for $n=10$ three-loop diagrams).
On the other hand, the final result for any of the Lorentz invariants occupies some KBytes only.
With access to improved hardware, we plan to push the calculation further to
$n=16$, cf. Ref.~\cite{Blumlein:2004xt}.

%
% ---------------------------------------------------------------------
%
\setcounter{equation}{0}
\section{Results}
\label{sec:results}
%
% ---------------------------------------------------------------------
%

Following the steps outlined above in Sec.~\ref{sec:calculation} we arrive at the 
results for the coefficient functions 
$C_{2, {\rm ns}}\,$, $C_{L,{\rm ns}}$ at the odd-integer values $n = 1,\, \ldots,\, 9,$ 
and for $C_{3,{\rm ns}}$ at the even-integer values $n = 2,\, \ldots ,\, 10,$
up to order $\alpha_{s}^3$.
The third order expressions represent new results of this article.
Using $a_{s}\equiv \alpha_{s}/(4 \pi)$ and the shorthand notation for the
$n$-th moment $C_{i,n}^{\rm ns} \equiv C_{i, {\rm ns}}(n)$ 
we find the following numerical values at the scale $\mu_r = \mu_f = Q$,
%
%------------------- Results in a form with floating point
%
\begin{eqnarray}
\label{eq:c2ns1}
  C_{2,1}^{\rm ns} & = &
         1
\, ,
\quad
\\
  C_{2,3}^{\rm ns} & = &
         1 
          + 3.222222222\, \ar          
       + \ar^2\,  (
           72.32720288
          - 11.125 \, \nf
          )
\\
& &\mbox{}
       +   \ar^3\,  (
           1948.031519
          - 496.5427343 \, \nf
          + 14.20173594 \, \nf^2
          )
         \nonumber\, ,
\quad
\\
 C_{2,5}^{\rm ns} & = &
         1 +  8.725925925\, \ar
       + \ar^2\, (
           220.4151827
          - 22.64048559 \, \nf
          )
\\
& &\mbox{}
       + \ar^3 \, (
          6925.814438
          - 1347.125829 \, \nf
          + 32.94421923 \, \nf^2
          )
 \nonumber\, ,
\quad
\\
 C_{2,7}^{\rm ns} & = &
        1   + 13.43677248\, \ar
          + \ar^2 \, (
           386.2911104
          - 33.10212484 \, \nf
          )
\\
& &\mbox{}
       + \ar^3 \, (
           13505.16600
          - 2298.472900 \, \nf
          + 52.34745652 \, \nf^2
          )
 \nonumber\, ,
\quad
\\
\label{eq:c2ns9}
 C_{2,9}^{\rm ns} & = &
        1+  17.47820105\, \ar 
         + \ar^2\,  (
            555.2720117
          - 42.50367619 \, \nf
          )
\\
& &\mbox{}
       + \ar^3\,  (
            20990.73668
          - 3278.689323 \, \nf
          + 71.31040423 \, \nf^2
          )
 \nonumber\, ,
\quad
\\
 C_{L,1}^{\rm ns} & = &
          + 2.666666666\, \ar 
                 + \ar ^2 \,  (
           61.33333333
          - 4.740740740 \, \nf
          )
\\
& &\mbox{}
       + \ar^3\,   (
           2313.911655
          - 405.2001359 \, \nf
          + 10.20576131 \, \nf^2
          )
 \nonumber\, ,
\quad
\\
 C_{L,3}^{\rm ns} & = &
         1.333333333\, \ar 
       + \ar^2\,   (
           52.40384466
          - 3.925925925 \, \nf
          )
\\
& &\mbox{}
       + \ar^3\,   (
           2584.178446
          - 406.0509532 \, \nf
          + 11.59670781 \, \nf^2
          )
 \nonumber\, ,
\quad
\\
 C_{L,5}^{\rm ns} & = &
            + 0.8888888888\, \ar 
       + \ar^2\,   (
           44.23466187
          - 3.012345679 \, \nf
          )
\\
& &\mbox{}
       + \ar^3\,   (
           2451.068575
          - 360.6487058 \, \nf
          + 10.15089163 \, \nf^2
          )
 \nonumber\, ,
\quad
\\
 C_{L,7}^{\rm ns} & = &
         0.6666666666\, \ar 
       + \ar^2\,   (
           38.25090234
          - 2.440740740 \, \nf
          )
\\
& &\mbox{}
       + \ar^3\,   (
           2290.679208
          - 321.7773285 \, \nf
          + 8.868930041 \, \nf^2
          ) 
 \nonumber\, ,
\quad
\\
 C_{L,9}^{\rm ns} & = &
        0.5333333333\,\ar 
       + \ar^2\,   (
          33.82305394
          - 2.056719576 \, \nf
          )
\\
& &\mbox{}
       + \ar^3\,   (
          2146.302724
          - 290.9906309 \, \nf
          + 7.868039976 \, \nf^2
          )
 \nonumber\, ,
\quad
\\
 C_{3,2}^{\rm ns} & = &
        1 - 1.777777777\, \ar 
       + \ar^2\,   (
          - 47.07704646
          - 0.09876543209 \, \nf
          )
\\
& &\mbox{}
       + \ar^3\,   (
          - 2359.001407
          + 305.2538856  \, \nf
          - 6.864103442 \, \nf^2
          )
 \nonumber\, ,
\quad
\\
 C_{3,4}^{\rm ns} & = &
        1           + 4.866666666\, \ar 
       + \ar^2\,   (
           90.15322509
          - 13.25902469 \, \nf
          )
\\
& &\mbox{}
       + \ar^3\,   (
           1478.747872
          - 491.0449098 \, \nf
          + 11.77903924 \, \nf^2
          )
 \nonumber\, ,
\quad
\\
 C_{3,6}^{\rm ns} & = &
        1 +  10.35132275\, \ar 
       + \ar^2\,   (
          258.8595696
          - 25.14210054 \, \nf
          )
\\
& &\mbox{}
       + \ar^3\,   (
           7586.717646
          - 1458.855783 \, \nf
          + 32.73965909 \, \nf^2
          )
 \nonumber\, ,
\quad
\\
 C_{3,8}^{\rm ns} & = &
        1 + 14.90026455\, \ar 
       + \ar^2\,   (
          433.2106396
          - 35.58166191 \, \nf
          )
\\
& &\mbox{}
       + \ar^3\,   (
          14862.60949
          - 2469.591886 \, \nf
          + 53.25812942 \, \nf^2
          )
 \nonumber\, ,
\quad
\\
\label{eq:c3ns10}
 C_{3,10}^{\rm ns} & = &
         1 + 18.79152477\, \ar 
       + \ar^2\,   (
          605.9424494
          - 44.87506803  \, \nf
          )
\\
& &\mbox{}
       + \ar^3\,    (
           22806.38215
          - 3482.933316 \, \nf
          + 72.83344233 \, \nf^2
          )
 \nonumber\, .
\end{eqnarray}
%
%------------------- End of results in a form with floating point
%
Exact analytical expressions of these moments also with complete dependence 
on the color coefficients are given in Appendix~\ref{sec:appA}.

As was mentioned before, the two-loop coefficient functions in
Eqs.~(\ref{eq:c2ns1})--(\ref{eq:c3ns10}) agree with 
the results in Refs.~\cite{vanNeerven:1991nn,Zijlstra:1991qc,Zijlstra:1992kj,Zijlstra:1992qd,Moch:1999eb}.
In addition, Eq.~(\ref{eq:c2ns1}) for $C_{2,1}^{\rm ns}$ is nothing else but
a manifestation of the Adler sum rule for DIS structure functions,
\begin{equation}
  \label{eq:adler-sumrule}
  \int\limits_0^1 \frac{dx}{x} \biggl(F_2^{\nu P}(x,Q^{2}) - F_2^{\nu N}(x,Q^{2}) \biggr) = 2 
\, ,
\end{equation}
which measures the isospin of the nucleon in the quark-parton model and does
not receive any perturbative or non-perturbative corrections in QCD, 
see e.g. the discussion in Ref.~\cite{Dokshitzer:1995qm}.
Therefore, Eq.~(\ref{eq:c2ns1}) is another important check of the correctness of our results.

\begin{figure}[ht]
  \begin{center}
    \includegraphics[width=8.15cm]{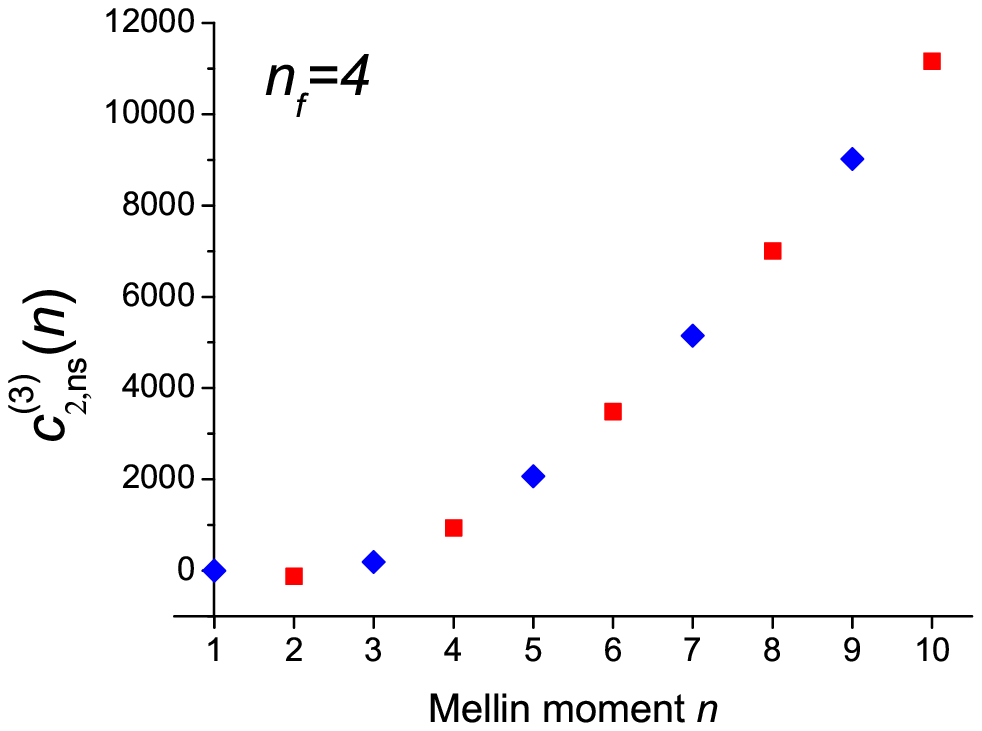}
    \includegraphics[width=8.15cm]{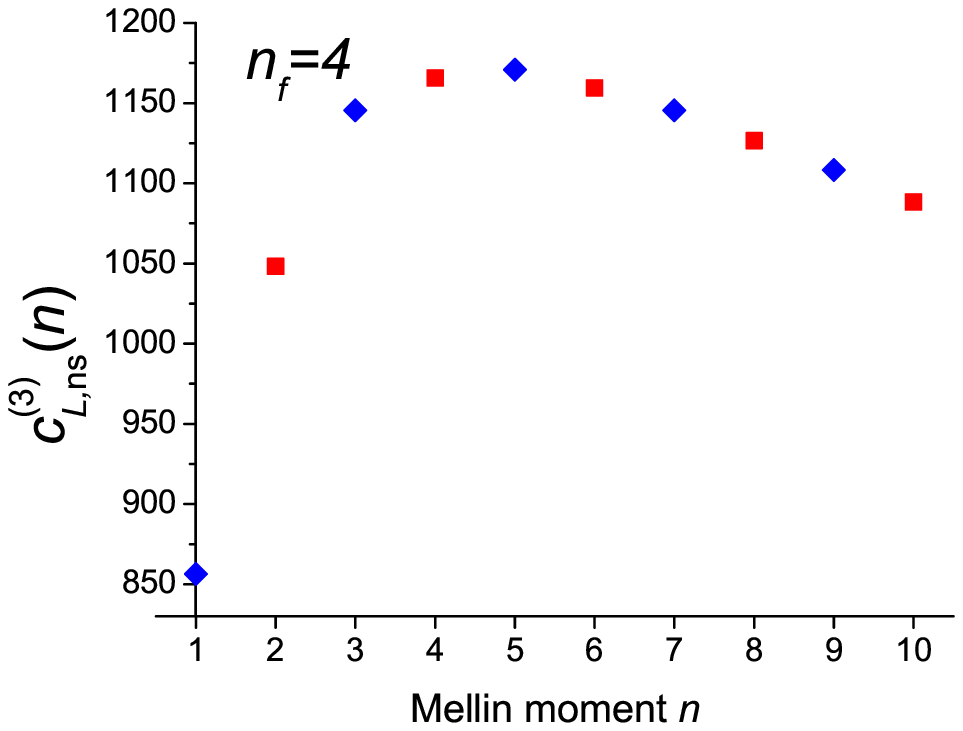}
    \caption[]{\label{fig:plotC2CL} 
The first ten integer Mellin moments of the third order non-singlet coefficient functions 
$c^{(3)}_{2,{\rm ns}}$ (left) and $c^{(3)}_{L,{\rm ns}}$ (right) for charged
current DIS with $n_f=4$ flavors. 
For the even moments, the flavor class $fl_{02}$ has been
omitted, i.e. $fl_{02}=0$ (see also the discussion in Sec.~\ref{sec:formalism}).
}
  \end{center}
\end{figure}
\begin{figure}[ht]
  \begin{center}
    \includegraphics[width=8.15cm]{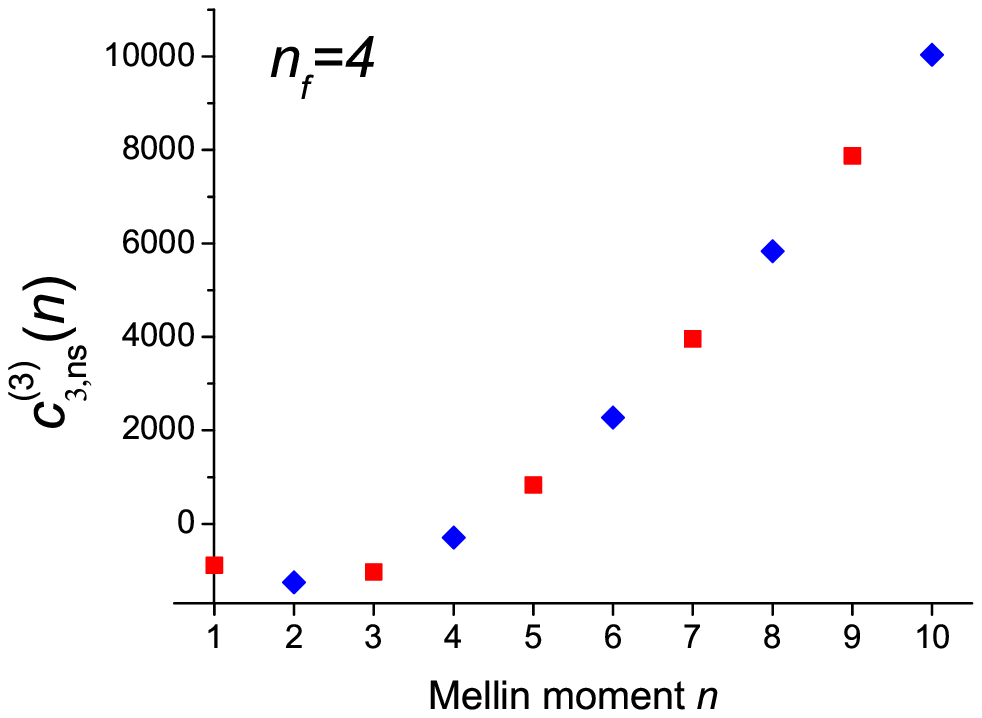}
    \caption[]{\label{fig:plotC3} 
The first ten integer Mellin moments of the third order non-singlet coefficient functions 
$c^{(3)}_{3,{\rm ns}}$ for charged current DIS  with $n_f=4$ flavors. 
For the odd moments, the flavor class $fl_{02}$ has been
omitted, i.e. $fl_{02}=0$ (see also the discussion in Sec.~\ref{sec:formalism}).
}
  \end{center}
\end{figure}
For illustration, let us plot the Mellin moments of the coefficient functions 
$c^{(3)}_{2,{\rm ns}}$, $c^{(3)}_{3,{\rm ns}}$ and $c^{(3)}_{L,{\rm ns}}$ at
three loops.
The new non-singlet results (blue diamonds in Figs.~\ref{fig:plotC2CL} and \ref{fig:plotC3})
exhibit a similar smooth pattern as the known results (red squares in Figs.~\ref{fig:plotC2CL} and \ref{fig:plotC3}).
Thus, it is feasible to use these moments for an approximate analytic reconstruction of the 
yet unknown coefficient functions $c^{(3)}_{i,{\rm ns}}$ for 
$(\nu P - {\bar \nu} P)$-DIS prior to ``all-$n$'' calculation and 
similar to e.g. Refs.~\cite{vanNeerven:2001pe,Vermaseren:2006ag}.
We will do so in a companion paper~\cite{MRV1}.

Furthermore, we see that the respective values for odd and even moments, 
for instance on the left in Fig.~\ref{fig:plotC2CL} do confirm that differences between 
$c^{(3)\, {\nu P + {\bar \nu} P}}_{2,{\rm ns}}$ and $c^{(3)\, {\nu P - {\bar \nu} P}}_{2,{\rm ns}}$ 
are numerically small. 
This observation (see Fig.~\ref{fig:plotC3}) provides also a posteriori justification for the
extrapolation procedure from odd to even moments for $C_3$ in Refs.~\cite{Kataev:1999bp,Kataev:2001kk}. 
There, available information on odd moments~\cite{Retey:2000nq} used in fits 
of CCFR data~\cite{Seligman:1997mc} to the structure function $xF_3$ at NNLO in QCD and beyond.
A further discussion of this and related issues is given in Ref.~\cite{MRV1}.

%
% ---------------------------------------------------------------------
%
\setcounter{equation}{0}
\section{Conclusions}
\label{sec:conclusions}
%
% ---------------------------------------------------------------------
%

In the present paper we have presented new results for Mellin moments 
of the charged current DIS structure functions $F_2^{\nu P - \bar \nu P}$, 
$F_L^{\nu P - \bar \nu P}$ and $F_3^{\nu P - \bar \nu P}$ including 
the perturbative QCD corrections to three loops.
In the former case ($F_2$, $F_L$) we have computed the first five odd-integer Mellin moments 
while in the latter case ($F_3$), the first five even-integer moments have been given.
Our efforts are part of an ongoing program~\cite{%
Moch:2004pa,Vogt:2004mw,vanNeerven:1991nn,Zijlstra:1991qc,Zijlstra:1992kj,Zijlstra:1992qd,%
Moch:1999eb,Moch:2004xu,Vermaseren:2005qc,Larin:1994vu,Larin:1997wd,Retey:2000nq,Moch:2001im,Blumlein:2004xt%
}
to calculate perturbative QCD corrections of all DIS structure functions to three-loop accuracy.

Within the framework of the OPE and the optical theorem we have calculated
Feynman diagrams of the forward Compton amplitude $T_{\mu\nu}$ in Mellin space.
In our presentation we have emphasized the symmetry properties of $T_{\mu\nu}$ 
and their relation to charged current ${\nu P \pm \bar \nu P}$ DIS, 
which was a crucial point in setting up the databases of Feynman diagrams.
We have performed various checks on our computation. 
Most prominently, we have kept all powers of the gauge parameter $\xi$ throughout 
the entire calculation to check that any $\xi$-dependence vanishes in our final results.
Furthermore, we agree with the literature as far as the 
two-loop coefficient functions~\cite{vanNeerven:1991nn,Zijlstra:1991qc,Zijlstra:1992kj,Zijlstra:1992qd,Moch:1999eb} 
and the three-loop anomalous dimensions~\cite{Moch:2004pa} are concerned.

The discussion of phenomenological consequences of our Mellin space results is deferred to Ref.~\cite{MRV1}.
Future research will be devoted to the calculation of some higher Mellin
moments, potentially $n=11,\dots,16$, depending on the available hardware infrastructure.
Subsequently, we will also focus on an ``all-$n$'' calculation in Mellin-$n$ space
with methods of Refs.~\cite{Moch:2004pa,Vogt:2004mw,Moch:2004xu,Vermaseren:2005qc},  
since all databases for Feynman diagrams contributing to 
$F_2^{ \nu P - \bar \nu P}$, $F_3^{ \nu P - \bar \nu P}$ and $F_L^{\nu P -\bar \nu P}$ 
are available now.

{\sc Form} files of these results can be obtained from the preprint server 
{\tt http://arXiv.org} by downloading the source of this article. 
Furthermore they are available from the authors upon request.

%
% ---------------------------------------------------------------------
%
{\bf{Acknowledgments:}}
We are grateful to J.~Vermaseren and A.~Vogt for useful discussions 
and to A.~Vogt for valuable comments on the manuscript.
The figures have been prepared with the packages {\sc Axodraw}~\cite{Vermaseren:1994je} and 
{\sc Jaxo\-draw}~\cite{Binosi:2003yf}. 
We acknowledge support by the Helmholtz Gemeinschaft under contract VH-NG-105 
and in part by the Deutsche Forschungsgemeinschaft in Sonderforschungs\-be\-reich/Transregio~9. 
%
% ---------------------------------------------------------------------
%

\appendix
%
% ---------------------------------------------------------------------
%
\renewcommand{\theequation}{\ref{sec:appA}.\arabic{equation}}
\setcounter{equation}{0}
\section{Appendix}
\label{sec:appA}
%
% ---------------------------------------------------------------------
%

In this Appendix we present the analytic expressions up to order $\ar^3$ 
for the coefficient functions 
$C_{2}^{\rm ns}$, $C_{L}^{\rm ns}$ at the odd-integer values $n = 1,\, \ldots,\, 9$ 
and for $C_{3}^{\rm ns}$ at the even-integer values $n = 2,\, \ldots ,\, 10$.
The notation follows Sec.~\ref{sec:results} with $C_A$ and $C_F$ 
being the standard QCD colour factors, 
$C_A \equiv N_c = 3$ and $C_F = (N_c^2 -1)/(2N_c) = 4/3$, and 
$\zeta_i$ stands for Riemann's $\zeta$-function. 
The coefficient functions for the structure function $F_2$ at
the scale $\mu_r = \mu_f = Q$ are given by
\begin{eqnarray}
\label{eq:c2q1}
  C_{2,1}^{\rm ns} & = &
%%START
%%L %%texc2q1 =
         1
%%;
%%STOP
\, ,
\quad
\\
\label{eq:c2q3}
  C_{2,3}^{\rm ns} & = &
%%START
%%L %%texc2q3 =
         1
\\
& &\mbox{}
       + \ar \* \cf \* {29 \over 12}
\nonumber\\
& &\mbox{}
       + \ar^2 \* \cf \* \nf  \*  \biggl(
          - {267 \over 32}
          \biggr)
\nonumber\\
& &\mbox{}
       + \ar^2 \* \cf^2  \*  \biggl(
          - {217235 \over 10368}
          + 36 \* \z3
          \biggr)
\nonumber\\
& &\mbox{}
       + \ar^2 \* \ca \* \cf  \*  \biggl(
            {25855 \over 432}
          - 43 \* \z3
          \biggr)
\nonumber\\
& &\mbox{}
       + \ar^3 \* \cf \* \nf^2  \*  \biggl(
            {641563 \over 69984}
          + {100 \over 81} \* \z3
          \biggr)
\nonumber\\
& &\mbox{}
       + \ar^3 \* \cf^2 \* \nf  \*  \biggl(
          - {35314337 \over 699840}
          - {1577 \over 45} \* \z3
          + {50 \over 3} \* \z4
          \biggr)
\nonumber\\
& &\mbox{}
       + \ar^3 \* \cf^3  \*  \biggl(
            {57093841 \over 373248}
          + {103235 \over 324} \* \z3
          + {55 \over 3} \* \z4
          - {1520 \over 3} \* \z5
          \biggr)
\nonumber\\
& &\mbox{}
       + \ar^3 \* \ca \* \cf \* \nf  \*  \biggl(
          - {132494393 \over 699840}
          + {39203 \over 405} \* \z3
          - {50 \over 3} \* \z4
          \biggr)
\nonumber\\
& &\mbox{}
       + \ar^3 \* \ca \* \cf^2  \*  \biggl(
          - {614328541 \over 2799360}
          + {16171 \over 45} \* \z3
          - {55 \over 2} \* \z4
          + 40 \* \z5
          \biggr)
\nonumber\\
& &\mbox{}
       + \ar^3 \* \ca^2 \* \cf  \*  \biggl(
            {490358569 \over 699840}
          - {344929 \over 405} \* \z3
          + {55 \over 6} \* \z4
          + {1070 \over 3} \* \z5
          \biggr)
%%;
%%STOP
\nonumber\, ,
\quad
\\
\label{eq:c2q5}
  C_{2,5}^{\rm ns} & = &
%%START
%%L %%texc2q5 =
         1
\\
& &\mbox{}
       + \ar \* \cf  \*  
          {589 \over 90}
\nonumber\\
& &\mbox{}
       + \ar^2 \* \cf \* \nf  \*  \biggl(
          - {2750819 \over 162000}
          \biggr)
\nonumber\\
& &\mbox{}
       + \ar^2 \* \cf^2  \*  \biggl(
          - {30297101 \over 1620000}
          + 52 \* \z3
          \biggr)
\nonumber\\
& &\mbox{}
       + \ar^2 \* \ca \* \cf  \*  \biggl(
            {35848409 \over 324000}
          - {312 \over 5} \* \z3
          \biggr)
\nonumber\\
& &\mbox{}
       + \ar^3 \* \cf \* \nf^2  \*  \biggl(
          {10958051 \over 486000}
          + {728 \over 405} \* \z3
          \biggr)
\nonumber\\
& &\mbox{}
       + \ar^3 \* \cf^2 \* \nf  \*  \biggl(
          - {38720716199 \over 255150000}
          - {257668 \over 4725} \* \z3
          + {364 \over 15} \* \z4
          \biggr)
\nonumber\\
& &\mbox{}
       + \ar^3 \* \cf^3  \*  \biggl(
           {132383443837 \over 2187000000}
          + {5617861 \over 10125} \* \z3
          + {1414 \over 75} \* \z4
          - 568 \* \z5
          \biggr)
\nonumber\\
& &\mbox{}
       + \ar^3 \* \ca \* \cf \* \nf  \*  \biggl(
          - {50351664421 \over 122472000}
          + {2187697 \over 14175} \* \z3
          - {364 \over 15} \* \z4
          \biggr)
 \nonumber\\
& &\mbox{}
      + \ar^3 \* \ca \* \cf^2  \*  \biggl(
          {9620298263 \over 63787500}
          + {10624153 \over 23625} \* \z3
          - {707 \over 25} \* \z4
          - 108 \* \z5
          \biggr)
\nonumber\\
& &\mbox{}
       + \ar^3 \* \ca^2 \* \cf  \*  \biggl(
          {230304417311 \over 163296000}
          - {16031641 \over 11340} \* \z3
          + {707 \over 75} \* \z4
          + 560 \* \z5
          \biggr)
%%;
%%STOP
\nonumber\, ,
\quad
\\
\label{eq:c2q7}
  C_{2,7}^{\rm ns} & = &
%%START
%%L %%texc2q7 =
       1 
\\
& &\mbox{}
       + \ar \* \cf  \*  
          {50791 \over 5040}
\nonumber\\
& &\mbox{}
       + \ar^2 \* \cf \* \nf  \*  \biggl(
          - {22072232029 \over 889056000}
          \biggr)
\nonumber\\
& &\mbox{}
       + \ar^2 \* \cf^2  \*  \biggl(
          - {430403824451 \over 248935680000}
          + {318 \over 5} \* \z3
          \biggr)
 \nonumber\\
& &\mbox{}
      + \ar^2 \* \ca \* \cf  \*  \biggl(
          {549422934719 \over 3556224000}
          - {5307 \over 70} \* \z3
          \biggr)
 \nonumber\\
& &\mbox{}
      + \ar^3 \* \cf \* \nf^2  \*  \biggl(
          { 12315998504291 \over 336063168000}
          +{ 2054 \over 945} \* \z3
          \biggr)
\nonumber\\
& &\mbox{}
       + \ar^3 \* \cf^2 \* \nf  \*  \biggl(
          {- 1296360189717461 \over 4356374400000}
          - {9299357 \over 132300} \* \z3
          + {1027 \over 35} \* \z4
         \biggr)
\nonumber\\
& &\mbox{}
       + \ar^3 \* \cf^3  \*  \biggl(
          - {1047772544741425169 \over 43912253952000000}
          + {31380651109 \over 37044000} \* \z3
          + {92741 \over 4900} \* \z4
          - {4388 \over 7} \* \z5
          \biggr)
\nonumber\\
& &\mbox{}
       + \ar^3 \* \ca \* \cf \* \nf  \*  \biggl(
          - {4190308663484983 \over 6721263360000}
          + {722354 \over 3675} \* \z3
          - {1027 \over 35} \* \z4
          \biggr)
\nonumber\\
& &\mbox{}
       + \ar^3 \* \ca \* \cf^2  \*  \biggl(
          {1465360668753075349 \over 1881953740800000}
           + {33630683 \over 82320} \* \z3
           - {278223 \over 9800} \* \z4
            - {1376 \over 7} \* \z5
           \biggr)
\nonumber\\
& &\mbox{}
       + \ar^3 \* \ca^2 \* \cf  \*  \biggl(
          {873626018834459 \over 420078960000}
          - {813517799 \over 441000} \* \z3
          + {92741 \over 9800} \* \z4
          + {4866 \over 7} \* \z5
          \biggr)
%%;
%%STOP
\nonumber\, ,
\quad
\\
\label{eq:c2q9}
  C_{2,9}^{\rm ns} & = &
%%START
%%L %%texc2q9 =
        1
\\
& &\mbox{}
       + \ar \* \cf  \* 
          {165169 \over 12600}
\nonumber\\
& &\mbox{}
       + \ar^2 \* \cf \* \nf  \*  \biggl(
          - {382605001967 \over 12002256000}
          \biggr)
\nonumber\\
& &\mbox{}
       + \ar^2 \* \cf^2  \*  \biggl(
          {4711040116777 \over 201637900800}
          + {510 \over 7} \* \z3
          \biggr)
\nonumber\\
& &\mbox{}
       + \ar^2 \* \ca \* \cf  \*  \biggl(
          {946052961283 \over 4898880000}
          - {1810 \over 21} \* \z3
          \biggr)
\nonumber\\
& &\mbox{}
       + \ar^3 \* \cf \* \nf^2  \*  \biggl(
          {3438632355495191 \over 68052791520000}
          + {4180 \over 1701} \* \z3
          \biggr)
 \nonumber\\
& &\mbox{}
      + \ar^3 \* \cf^2 \* \nf  \*  \biggl(
          - {73610396284048043863 \over 157201948411200000}
          - {281314024 \over 3274425} \* \z3
          + {2090 \over 63} \* \z4
          \biggr)
\nonumber\\
& &\mbox{}
       + \ar^3 \* \cf^3  \*  \biggl(
          - {202884354298201249627 \over 4001504141376000000}
          + {110079158608 \over 93767625} \* \z3
          + {1253219 \over 66150} \* \z4
          - 708 \* \z5
          \biggr)
\nonumber\\
& &\mbox{}
       + \ar^3 \* \ca \* \cf \* \nf  \*  \biggl(
          - {31467396414071567 \over 38192893200000}
          + {1515609253 \over 6548850} \* \z3
          - {2090 \over 63} \* \z4
          \biggr)
 \nonumber\\
& &\mbox{}
      + \ar^3 \* \ca \* \cf^2  \*  \biggl(
          {883676953019117176289 \over 571643448768000000}
          + {6226321733 \over 20837250} \* \z3
          - {1253219 \over 44100} \* \z4
          - {2078 \over 9} \* \z5
          \biggr)
\nonumber\\
& &\mbox{}
       + \ar^3 \* \ca^2 \* \cf  \*  \biggl(
          {73714919752951175633 \over 27221116608000000}
           - {872558077 \over 396900} \* \z3
           + {1253219 \over 132300} \* \z4
          + {49774 \over 63} \* \z5
          \biggr)
%%;
%%STOP
\nonumber\, .
\end{eqnarray}

The coefficient functions for the structure function $F_L$ at
the scale $\mu_r = \mu_f = Q$ are given by
\begin{eqnarray}
\label{eq:cLq1}
  C_{L,1}^{\rm ns} & = &
%%START
%%L %%texcLq1 =
       \ar \* \cf  \*  
          2
\\
& &\mbox{}
       + \ar^2 \* \cf \* \nf  \*  \biggl(
          - {32 \over 9}
          \biggr)
\nonumber\\
& &\mbox{}
       + \ar^2 \* \cf^2  \*  \biggl(
          - 11
          \biggr)
\nonumber\\
& &\mbox{}
       + \ar^2 \* \ca \* \cf  \*  
          {182 \over 9}
\nonumber\\
& &\mbox{}
       + \ar^3 \* \cf \* \nf^2  \* 
           {620 \over 81}
 \nonumber\\
& &\mbox{}
      + \ar^3 \* \cf^2 \* \nf  \*  \biggl(
          { 335 \over 9}
          - {16 \over 3} \* \z3
          \biggr)
\nonumber\\
& &\mbox{}
       + \ar^3 \* \cf^3  \*  \biggl(
          313
           + 752 \* \z3
          - 1120 \* \z5
           \biggr)
 \nonumber\\
& &\mbox{}
      + \ar^3 \* \ca \* \cf \* \nf  \*  \biggl(
          - {8470 \over 81}
          + {112 \over 3} \* \z3
          - {160 \over 3} \* \z5
          \biggr)
\nonumber\\
& &\mbox{}
       + \ar^3 \* \ca \* \cf^2  \*  \biggl(
          - {5462 \over 9}
          - {2912 \over 3}\* \z3
          + 1520 \* \z5
          \biggr)
 \nonumber\\
& &\mbox{}
      + \ar^3 \* \ca^2 \* \cf  \*  \biggl(
          {33140 \over 81}
          + 160 \* \z3
          - 320 \* \z5
          \biggr)
%%;
%%STOP
\nonumber\, ,
\quad
\\
\label{eq:cLq3}
  C_{L,3}^{\rm ns} & = &
%%START
%%L %%texcLq3 =
      \ar \* \cf           
\\
& &\mbox{}
      + \ar^2 \* \cf \* \nf  \*  \biggl(
          - {53 \over 18}
          \biggr)
 \nonumber\\
& &\mbox{}
      + \ar^2 \* \cf^2  \*  \biggl(
          - {607 \over 24}
          + 24 \* \z3
          \biggr)
  \nonumber\\
& &\mbox{}
     + \ar^2 \* \ca \* \cf  \*  \biggl(
           {467 \over 18}
          - 12 \* \z3
          \biggr)
\nonumber\\
& &\mbox{}
       + \ar^3 \* \cf \* \nf^2  \* 
          { 1409 \over 162}
 \nonumber\\
& &\mbox{}
      + \ar^3 \* \cf^2 \* \nf  \*  \biggl(
          {403511 \over 4320}
          - {1496 \over 15} \* \z3
          \biggr)
 \nonumber\\
& &\mbox{}
      + \ar^3 \* \cf^3  \*  \biggl(
          {2898181 \over 10368}
          + {2434 \over 9 }\* \z3
          - 560 \* \z5
          \biggr)
 \nonumber\\
& &\mbox{}
      + \ar^3 \* \ca \* \cf \* \nf  \*  \biggl(
          - {253051 \over 1620}
          + {2488 \over 45} \* \z3
          \biggr)
 \nonumber\\
& &\mbox{}
      + \ar^3 \* \ca \* \cf^2  \*  \biggl(
          - {979877 \over 1080}
          + {22184 \over 45} \* \z3
          + 360 \* \z5
          \biggr)
 \nonumber\\
& &\mbox{}
      + \ar^3 \* \ca^2 \* \cf  \*  \biggl(
           {4262777 \over 6480}
          - {1796 \over 5} \* \z3
          - 40 \* \z5
          \biggr)
%%;
%%STOP
\nonumber\, ,
\quad
\\
\label{eq:cLq5}
  C_{L,5}^{\rm ns} & = &
%%START
%%L %%texcLq5 =
        \ar \* \cf  \*  
          {2 \over 3}
\\
& &\mbox{}
       + \ar^2 \* \cf \* \nf  \*  \biggl(
          - {61 \over 27}
          \biggr)
\nonumber\\
& &\mbox{}
       + \ar^2 \* \cf^2  \*  \biggl(
          - {119 \over 9}
          + 16 \* \z3
          \biggr)
 \nonumber\\
& &\mbox{}
      + \ar^2 \* \ca \* \cf  \*  \biggl(
          {4861 \over 270}
          - 8 \* \z3
          \biggr)
\nonumber\\
& &\mbox{}
       + \ar^3 \* \cf \* \nf^2  \* 
          {1850 \over 243}
\nonumber\\
& &\mbox{}
       + \ar^3 \* \cf^2 \* \nf  \*  \biggl(
          {5479121 \over 63000}
          - {11128 \over 105} \* \z3
          \biggr)
 \nonumber\\
& &\mbox{}
      + \ar^3 \* \cf^3  \*  \biggl(
          {107779259 \over 2430000}
          + {320 \over 3 } \* \z5
          - {64426 \over 675} \* \z3
          \biggr)
\nonumber\\
& &\mbox{}
       + \ar^3 \* \ca \* \cf \* \nf  \*  \biggl(
          - {48931997 \over 340200}
          +{ 56318 \over 945 } \* \z3
          \biggr)
\nonumber\\
& &\mbox{}
       + \ar^3 \* \ca \* \cf^2  \*  \biggl(
          - {73036627 \over 113400}
          + {3980216 \over 4725} \* \z3
          - 240 \* \z5
          \biggr)
 \nonumber\\
& &\mbox{}
      + \ar^3 \* \ca^2 \* \cf  \*  \biggl(
          {784906033 \over 1360800}
          - {278009 \over 630} \* \z3
          + {280 \over 3 }\* \z5
          \biggr)
%%;
%%STOP
\nonumber\, ,
\quad
\\
\label{eq:cLq7}
  C_{L,7}^{\rm ns} & = &
%%START
%%L %%texcLq7 =  
    \ar \* \cf  \* 
          {1 \over 2}
\\
& &\mbox{}
      + \ar^2 \* \cf \* \nf  \*  \biggl(
          - {659 \over 360}
          \biggr)
\nonumber\\
& &\mbox{}
       + \ar^2 \* \cf^2  \*  \biggl(
          - {2351887 \over 302400}
          + 12 \* \z3
          \biggr)
\nonumber\\
& &\mbox{}
       + \ar^2 \* \ca \* \cf  \*  \biggl(
          {2089693 \over 151200}
          - 6 \* \z3
          \biggr)
 \nonumber\\
& &\mbox{}
      + \ar^3 \* \cf \* \nf^2  \* 
          {43103 \over 6480}
\nonumber\\
& &\mbox{}
       + \ar^3 \* \cf^2 \* \nf  \*  \biggl(
          {1089629735311 \over 16003008000}
          - {148192 \over 1575} \* \z3
          \biggr)
 \nonumber\\
& &\mbox{}
      + \ar^3 \* \cf^3  \*  \biggl(
          - {10999391897239 \over 99574272000}
          + 600 \* \z5
          - {83782639 \over 220500} \* \z3
          \biggr)
\nonumber\\
& &\mbox{}
       + \ar^3 \* \ca \* \cf \* \nf  \*  \biggl(
          - {7151831837 \over 57153600}
          + {10172 \over 189} \* \z3
          \biggr)
 \nonumber\\
& &\mbox{}
      + \ar^3 \* \ca \* \cf^2  \*  \biggl(
          - {17816756739419 \over 45722880000}
          + {460935289 \over 441000 } \* \z3
          - 700 \* \z5
          \biggr)
 \nonumber\\
& &\mbox{}
      + \ar^3 \* \ca^2 \* \cf  \*  \biggl(
          {1082520395023 \over 2286144000}
          - {14765939 \over 31500} \* \z3
          + 200 \* \z5
          \biggr)
%%;
%%STOP
\nonumber\, ,
\quad
\\
\label{eq:cLq9}
  C_{L,9}^{\rm ns} & = &
%%START
%%L %%texcLq9 =
        \ar \* \cf  \*  
          {2 \over 5}
\\
& &\mbox{}
       + \ar^2 \* \cf \* \nf  \*  \biggl(
          - {4859 \over 3150}
          \biggr)
\nonumber\\
& &\mbox{}
       + \ar^2 \* \cf^2  \*  \biggl(
          - {3177697 \over 661500}
          + {48 \over 5 } \* \z3
          \biggr)
\nonumber\\
& &\mbox{}
       + \ar^2 \* \ca \* \cf  \*  \biggl(
          {7429883 \over 661500}
          - {24 \over 5} \* \z3
          \biggr)
 \nonumber\\
& &\mbox{}
      + \ar^3 \* \cf \* \nf^2  \* 
          {836471 \over 141750}
  \nonumber\\
& &\mbox{}
     + \ar^3 \* \cf^2 \* \nf  \*  \biggl(
          {17671408832087 \over 330062040000}
          - {1439512 \over 17325} \* \z3
          \biggr)
 \nonumber\\
& &\mbox{}
      + \ar^3 \* \cf^3  \*  \biggl(
          - {5753430631305541 \over 25204737600000}
          - {35201422 \over 55125} \* \z3
          + 1024 \* \z5
          \biggr)
 \nonumber\\
& &\mbox{}
      + \ar^3 \* \ca \* \cf \* \nf  \*  \biggl(
          - {1186239473563 \over 10777536000}
          + {17528393 \over 363825} \* \z3
          \biggr)
  \nonumber\\
& &\mbox{}
     + \ar^3 \* \ca \* \cf^2  \*  \biggl(
          - {174689016402059 \over 933508800000}
          + {45254429 \over 36750} \* \z3
          - 1104 \* \z5
          \biggr)
 \nonumber\\
& &\mbox{}
      + \ar^3 \* \ca^2 \* \cf  \*  \biggl(
          {88521637399093 \over 228614400000}
           - {41525 \over 84} \* \z3
          + 296 \* \z5
          \biggr)
%%;
%%STOP
\nonumber\, .
\end{eqnarray}

The coefficient functions for the structure function $F_3$ at
the scale $\mu_r = \mu_f = Q$ are given by
\begin{eqnarray}
\label{eq:c3q2}
  C_{3,2}^{\rm ns} & = &
%%START
%%L %%texc3q2 =
            1
\\
& &\mbox{}
    + \ar \* \cf  \*  \biggl(
          - {4 \over 3}
          \biggr)
\nonumber\\
& &\mbox{}
       + \ar^2 \* \cf \* \nf  \*  \biggl(
          - {2 \over 27}
          \biggr)
  \nonumber\\
& &\mbox{}
     + \ar^2 \* \cf^2  \*  
          {1016 \over 81}
 \nonumber\\
& &\mbox{}
      + \ar^2 \* \ca \* \cf  \*  \biggl(
          {17 \over 9}
          - 16 \* \z3
          \biggr)
 \nonumber\\
& &\mbox{}
      + \ar^3 \* \cf \* \nf^2  \*  \biggl(
          - {13336 \over 2187}
          + {64 \over 81} \* \z3
          \biggr)
  \nonumber\\
& &\mbox{}
     + \ar^3 \* \cf^2 \* \nf  \*  \biggl(
          - {182014 \over 2187}
          + {448 \over 9}\* \z3
          + {32 \over 3} \* \z4
          \biggr)
 \nonumber\\
& &\mbox{}
      + \ar^3 \* \cf^3  \*  \biggl(
          - {55954 \over 729}
          - {11584 \over 81} \* \z3
          + {64 \over 3} \* \z4
          + {640 \over 3} \* \z5
          \biggr)
 \nonumber\\
& &\mbox{}
      + \ar^3 \* \ca \* \cf \* \nf  \*  \biggl(
          {156404 \over 2187}
          + {1456 \over 81}\* \z3
          - {32 \over 3} \* \z4
          \biggr)
 \nonumber\\
& &\mbox{}
      + \ar^3 \* \ca \* \cf^2  \*  \biggl(
          {855382 \over 2187}
          + {56 \over 9} \* \z3
          - 32 \* \z4
          - 320 \* \z5
          \biggr)
 \nonumber\\
& &\mbox{}
      + \ar^3 \* \ca^2 \* \cf  \*  \biggl(
          - {481450 \over 2187}
          - {18728 \over 81} \* \z3
          + {32 \over 3 } \* \z4
          + {800 \over 3} \* \z5
          \biggr)
%%;
%%STOP
\nonumber\, ,
\quad
\\
\label{eq:c3q4}
  C_{3,4}^{\rm ns} & = &
%%START
%%L %%texc3q4 =
          1
\\
& &\mbox{}
       + \ar \* \cf  \*  
          {73 \over 20}
\nonumber\\
& &\mbox{}
       + \ar^2 \* \cf \* \nf  \*  \biggl(
          -{ 1073981 \over 108000}
          \biggr)
 \nonumber\\
& &\mbox{}
      + \ar^2 \* \cf^2  \*  \biggl(
          - {59219099 \over 6480000}
          + 28 \* \z3
          \biggr)
 \nonumber\\
& &\mbox{}
      + \ar^2 \* \ca \* \cf  \*  \biggl(
          {3575579 \over 54000}
          - {227 \over 5 } \* \z3
          \biggr)
 \nonumber\\
& &\mbox{}
      + \ar^3 \* \cf \* \nf^2  \*  \biggl(
          {12195323 \over 1749600}
          + {628 \over 405 } \* \z3
          \biggr)
    \nonumber\\
& &\mbox{}
   + \ar^3 \* \cf^2 \* \nf  \*  \biggl(
          - {18625311191 \over 109350000}
          + {38021 \over 675} \* \z3
          + {314 \over 15} \* \z4
          \biggr)
 \nonumber\\
& &\mbox{}
      + \ar^3 \* \cf^3  \*  \biggl(
          - {48030418393 \over 5832000000}
          + {9183239 \over 40500} \* \z3
          + {1439 \over 75} \* \z4
          - {704 \over 3} \* \z5
          \biggr)
\nonumber\\
& &\mbox{}
       + \ar^3 \* \ca \* \cf \* \nf  \*  \biggl(
          - {529878917 \over 3499200}
          + {29266 \over 405} \* \z3
          - {314 \over 15} \* \z4
          \biggr)
\nonumber\\
& &\mbox{}
       + \ar^3 \* \ca \* \cf^2  \*  \biggl(
          {1003904196083 \over 1749600000}
          - {185929 \over 2250} \* \z3
          - {1439 \over 50} \* \z4
          - 208 \* \z5
          \biggr)
\nonumber\\
& &\mbox{}
       + \ar^3 \* \ca^2 \* \cf  \*  \biggl(
          {8293616147 \over 17496000}
          - {1625431 \over 2025} \* \z3
          + {1439 \over 150} \* \z4
          + {1430 \over 3} \* \z5
          \biggr)
%%;
%%STOP
\nonumber\, ,
\quad
\\
\label{eq:c3q6}
  C_{3,6}^{\rm ns} & = &
%%START
%%L %%texc3q6 =
            1
\\
& &\mbox{}
    + \ar \* \cf  \* 
          {4891 \over 630}
 \nonumber\\
& &\mbox{}
      + \ar^2 \* \cf \* \nf  \*  \biggl(
          - {1047784469 \over 55566000}
          \biggr)
\nonumber\\
& &\mbox{}
       + \ar^2 \* \cf^2  \*  \biggl(
          - {15792028349 \over 3889620000}
          + {228 \over 5} \* \z3
          \biggr)
\nonumber\\
& &\mbox{}
       + \ar^2 \* \ca \* \cf  \*  \biggl(
          {105144247 \over 889056}
          - {2216 \over 35} \* \z3
          \biggr)
\nonumber\\
& &\mbox{}
       + \ar^3 \* \cf \* \nf^2  \*  \biggl(
          {25846271107 \over 1166886000}
          + {5672 \over 2835} \* \z3
          \biggr)
\nonumber\\
& &\mbox{}
       + \ar^3 \* \cf^2 \* \nf  \*  \biggl(
          - {57923821071217 \over 204205050000}
          + {212692 \over 6615} \* \z3
          + {2836 \over 105} \* \z4
          \biggr)
\nonumber\\
& &\mbox{}
       + \ar^3 \* \cf^3  \*  \biggl(
          - {10620928547301491 \over 257298363000000}
          + {2025255523 \over 3472875} \* \z3
          + {69862 \over 3675} \* \z4
          - {9944 \over 21} \* \z5
          \biggr)
\nonumber\\
& &\mbox{}
       + \ar^3 \* \ca \* \cf \* \nf  \*  \biggl(
          - {81497704436303 \over 210039480000}
          + {1750321 \over 14175} \* \z3
          - {2836 \over 105} \* \z4
          \biggr)
 \nonumber\\
& &\mbox{}
      + \ar^3 \* \ca \* \cf^2  \*  \biggl(
          {8198632169233 \over 8508543750}
          - {4570231 \over 42875} \* \z3
          - {34931 \over 1225} \* \z4
          - {1236 \over 7} \* \z5
          \biggr)
\nonumber\\
& &\mbox{}
       + \ar^3 \* \ca^2 \* \cf  \*  \biggl(
          {345351668819093 \over 280052640000}
          - {14709577 \over 11340} \* \z3
          + {34931 \over 3675} \* \z4
          + {12848 \over 21} \* \z5
          \biggr)
%%;
%%STOP
\nonumber\, ,
\quad
\\
\label{eq:c3q8}
  C_{3,8}^{\rm ns} & = &
%%START
%%L %%texc3q8 =
         1
\\
& &\mbox{}
       + \ar \* \cf  \* 
          {56323 \over 5040}
\nonumber\\
& &\mbox{}
       + \ar^2 \* \cf \* \nf  \*  \biggl(
          - {640590322783 \over 24004512000}
          \biggr)
\nonumber\\
& &\mbox{}
       + \ar^2 \* \cf^2  \*  \biggl(
          {292667334922909 \over 20163790080000}
          + {2046 \over 35} \* \z3
          \biggr)
\nonumber\\
& &\mbox{}
       + \ar^2 \* \ca \* \cf  \*  \biggl(
          {124690328633 \over 768144384}
          - {16021 \over 210} \* \z3
          \biggr)
\nonumber\\
& &\mbox{}
       + \ar^3 \* \cf \* \nf^2  \*  \biggl(
          {1011263478114371 \over 27221116608000}
          + {19766 \over 8505} \* \z3
          \biggr)
\nonumber\\
& &\mbox{}
       + \ar^3 \* \cf^2 \* \nf  \*  \biggl(
          - {6152314125227207753 \over 14291086219200000}
          + {1470167 \over 238140} \* \z3
          + {9883 \over 315} \* \z4
          \biggr)
\nonumber\\
& &\mbox{}
       + \ar^3 \* \cf^3  \*  \biggl(
          - {2725949924349586498181 \over 32012033131008000000}
          + {2676254422397 \over 3000564000} \* \z3
          + {2510407 \over 132300} \* \z4
          - 564 \* \z5
          \biggr)
\nonumber\\
& &\mbox{}
       + \ar^3 \* \ca \* \cf \* \nf  \*  \biggl(
          - { 332470249596994151 \over 544422332160000}
          + {198295567 \over 1190700} \* \z3 
          - {9883 \over 315} \* \z4
          \biggr)
\nonumber\\
& &\mbox{}
       + \ar^3 \* \ca \* \cf^2  \*  \biggl(
          {724070773013223516571 \over 457314759014400000}
          - {19725946973 \over 166698000} \* \z3
          - {2510407 \over 88200} \* \z4
          - {2132 \over 9} \* \z5
           \biggr)
\nonumber\\
& &\mbox{}
       + \ar^3 \* \ca^2 \* \cf  \*  \biggl(
          {528130813453946861 \over 272211166080000}
          - {39055999 \over 22680} \* \z3
          + {2510407 \over 264600} \* \z4
          + {45994 \over 63} \* \z5
          \biggr)
%%;
%%STOP
\nonumber\, ,
\quad
\\
\label{eq:c3q10}
  C_{3,10}^{\rm ns} & = &
%%START
%%L %%texc3q10 =
         1
\\
& &\mbox{}
       + \ar \* \cf  \*  
          {1953379 \over 138600}
\nonumber\\
& &\mbox{}
       + \ar^2 \* \cf \* \nf  \*  \biggl(
          - {537659500957277 \over 15975002736000}
          \biggr)
\nonumber\\
& &\mbox{}
       + \ar^2 \* \cf^2  \*  \biggl(
          {597399446375524589 \over 14760902528064000}
          + {7202 \over 105} \* \z3
          \biggr)
\nonumber\\
& &\mbox{}
       + \ar^2 \* \ca \* \cf  \*  \biggl(
          {5832602058122267 \over 29045459520000}
          - {99886 \over 1155} \* \z3
          \biggr)
\nonumber\\
& &\mbox{}
       + \ar^3 \* \cf \* \nf^2  \*  \biggl(
          {51339756673194617191 \over 996360920644320000}
          + {48220 \over 18711} \* \z3
          \biggr)
\nonumber\\
& &\mbox{}
       + \ar^3 \* \cf^2 \* \nf  \*  \biggl(
          - {125483817946055121351353 \over 209235793335307200000}
           - {59829376 \over 3274425} \* \z3
          + {24110 \over 693} \* \z4
          \biggr)
\nonumber\\
& &\mbox{}
       + \ar^3 \* \cf^3  \*  \biggl(
          - {744474223606695878525401307 \over 7088908678200207936000000}
          + {28630985464358 \over 24960941775} \* \z3
 \nonumber\\
& &\mbox{}\quad\quad\quad
          + {151796299 \over 8004150} \* \z4
          - {53708 \over 99} \* \z5
          \biggr)
\nonumber\\
& &\mbox{}
       + \ar^3 \* \ca \* \cf \* \nf  \*  \biggl(
          - {185221350045507487753 \over 226445663782800000}
          + {8071097 \over 39690} \* \z3
          - {24110 \over 693} \* \z4
          \biggr)
 \nonumber\\
& &\mbox{}
      + \ar^3 \* \ca \* \cf^2  \*  \biggl(
          {19770078729338607732075449 \over 8369431733412288000000}
          - {619383700181 \over 5546875950} \* \z3
 \nonumber\\
& &\mbox{}\quad\quad\quad
          - {151796299 \over 5336100} \* \z4
          - {37322 \over 99} \* \z5
          \biggr)
\nonumber\\
& &\mbox{}
       + \ar^3 \* \ca^2 \* \cf  \*  \biggl(
          {93798719639056648125143 \over 36231306205248000000}
          - {43202630363 \over 20582100} \* \z3
 \nonumber\\
& &\mbox{}\quad\quad\quad
          + {151796299 \over 16008300} \* \z4
          + {195422 \over 231} \* \z5
          \biggr)
%%;
%%STOP
\nonumber\, .
\end{eqnarray}
%

%
% ---------------------------------------------------------------------
%
\renewcommand{\theequation}{\ref{sec:appB}.\arabic{equation}}
\setcounter{equation}{0}
\section{Appendix}
\label{sec:appB}
%
% ---------------------------------------------------------------------
%

In this Appendix, we recall a few technical steps necessary to
arrive at the relations~(\ref{eq:F2mellinMR}), (\ref{eq:F3mellinMR}) between 
the parameters of OPE and the Mellin moments of DIS structure functions. 
To that end, we would like to put particular emphasis on the symmetry
properties of the hadron forward Compton amplitude $T_{\mu\nu}(p,q)$ in Eq.~(\ref{eq:forwardcompton}) 
under the transformations $\mu\leftrightarrow\nu$ and $q\rightarrow -q$. 
Since our discussion in Sec.~\ref{sec:formalism} was largely based on Feynman 
diagram considerations at parton level it remains to link the line of arguments to the OPE of
Eq.~(\ref{eq:OPE}), to the analogue of the OPE~(\ref{eq:OPEpartons}) for $T_{\mu\nu}$ and, eventually to the Mellin
moments of $F_2$, $F_3$ and $F_L$ in Eqs.~(\ref{eq:mellindefF2L}), (\ref{eq:mellindefF3}).

Let us start by observing, that the OPE~(\ref{eq:OPE}) for $T_{\mu\nu}$ gives
rise to a series expansion in terms of $\omega$ similar to Eq.~(\ref{eq:OPEpartons}).
This series being valid for unphysical $\omega=1/x\rightarrow 0$ only 
(recall the Bjorken variable $0<x \leq1$) is related to the physical Mellin
moments of $F_2$, $F_3$ and $F_L$ by means of a Cauchy integration.
Here, the behavior of $T_{\mu\nu}$ under the mapping $\omega \rightarrow -\omega$ 
($q\rightarrow-q$) becomes relevant.
Applying the Lorentz projectors~(\ref{eq:projL})--(\ref{eq:proj3}) to $T_{\mu\nu}$ we obtain 
\begin{eqnarray}
\label{eq:invarTmunu2L}
T_i(\omega,Q^{2})\equiv P_{L}^{\mu\nu}T_{\mu\nu}(\omega, Q^{2})=2 \sum_{n,j} {\omega}^n  C_{L,j}\left(n,\frac{Q^2}{\m^2},\a_s\right) 
 A_{{\rm{nucl}}}^{j}\left(n,{\m^2}\right)
\, , 
\quad\quad\quad
i = 2,L
\, .
\end{eqnarray}
In Sec.~\ref{sec:formalism} we discussed which coefficients $C_{L,j}\left(n\right)$ survive in the OPE~(\ref{eq:OPE}).
As consequence the sum~(\ref{eq:invarTmunu2L}) runs over even $n$ for the neutral current structure functions $F_2$, $F_L$ 
and the charged current (singlet) structure functions $F_2^{\nu P + \nu N}$, $F_L^{\nu P + \nu N}$ 
and therefore $T_i(-\omega, Q^{2})=T_i(\omega, Q^{2})$. 
For the charged current (non-singlet) structure functions 
$F_2^{\nu P - \nu N}$ and $F_L^{\nu P - \nu N}$ 
we sum over all odd $n$. Therefore $T_i(-\omega, Q^{2})=-T_i(\omega, Q^{2})$. 
Furthermore, we have
\begin{eqnarray}
\label{eq:invarTmunu3}
T_{3}(\omega,Q^{2})\equiv P_{3}^{\mu\nu}T_{\mu\nu}(\omega, Q^{2})=2 \sum_{n,j} {\omega}^n C_{3,j}\left(n,\frac{Q^2}{\m^2},\a_s\right)
A_{{\rm{nucl}}}^{j}\left(n,\mu^{2} \right)\, ,  
\end{eqnarray}
where we sum over all odd $n$ for the neutral current $F_3$ and charged current 
$F_{3}^{\nu P + \nu N}$. As consequence we have obviously $T_{3}(-\omega, Q^{2})=-T_{3}(\omega, Q^{2})$. 
On the other hand, for the charged current $F_{3}^{\nu P - \nu N}$ the sum runs 
over even $n$ and therefore $T_{3}(-\omega, Q^{2})=T_{3}(\omega, Q^{2})$.

\begin{figure}[ht]
  \begin{center}
    \includegraphics[width=6.0cm]{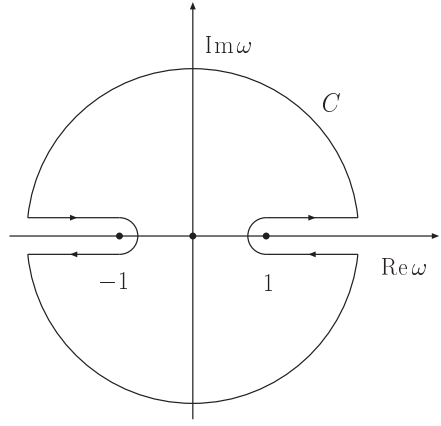}
    \caption[]{\label{fig:Cauchy} The contour $C$ of the Cauchy integration in the complex $\omega$-plane.}
  \end{center}
\end{figure}
In applying the Cauchy integration to both sides of Eqs.~(\ref{eq:invarTmunu2L}), (\ref{eq:invarTmunu3}) 
with a contour $C$ as shown in the Fig.~\ref{fig:Cauchy} 
we exploit the fact that the Lorentz projected forward Compton amplitude is an analytic function 
of complex variable $\omega$. 
The branch cuts extend along the real axis for $\omega \leq -1$ and $\omega \geq 1$ 
because of kinematical constraints from Bjorken $x$ and symmetry properties. We divide both sides of  Eqs.~(\ref{eq:invarTmunu2L}), (\ref{eq:invarTmunu3}) by $2\pi i \omega^{m}$
and pick up the appropriate residues on the r.h.s according to 
\begin{eqnarray}
\label{eq:Cauchy_omega}
\frac{1}{2\pi{\rm i}}\oint\limits_{C} d\omega \frac{\omega^{n}}{\omega^{m}}=\delta_{n,m-1}
\, .  
\end{eqnarray}
For the l.h.s. one obtains
\begin{eqnarray}
\label{eq:CauchyTmunu}
\frac{1}{2 \pi{\rm i}}\oint\limits_{C} \frac{d \omega}{\omega^{m}}\,T_i(\omega, Q^{2}) 
&=&
\frac{1}{2 \pi{\rm i}}\biggl(
+ \int\limits_{+\infty-{\rm i}\epsilon}^{+1-{\rm i}\epsilon}
+\int\limits_{+1+{\rm i}\epsilon}^{+\infty+{\rm i}\epsilon}
+\int\limits_{-\infty+{\rm i}\epsilon}^{-1+{\rm i}\epsilon}
+\int\limits_{-1-{\rm i}\epsilon}^{-\infty-{\rm i}\epsilon}\,\,
\biggr) \frac{d \omega}{\omega^{m}}\, T_i(\omega, Q^{2})
\\
&=&
\frac{1}{ \pi{\rm i}}\biggl(
- \int\limits^{+\infty-{\rm i}\epsilon}_{+1-{\rm i}\epsilon}
+\int\limits_{+1+{\rm i}\epsilon}^{+\infty+{\rm i}\epsilon}\,\,
\biggr) \frac{d \omega}{\omega^{m}}\, T_i(\omega, Q^{2})
\nonumber\\
&=&
\frac{2}{\pi }\int\limits_{1}^{+\infty}\frac{d \omega}{\omega^m}\, {\rm Im} T_i(\omega,{Q^2})
\, , 
\quad\quad\quad
i = 2,3,L
\, ,
\nonumber
\end{eqnarray}
where we have used that for the physical cases with either even or odd $n$
in Eqs.~(\ref{eq:invarTmunu2L}), (\ref{eq:invarTmunu3}) 
the whole combination $({d \omega}/{\omega^m})\, T_i(\omega, Q^{2})$ 
occurs to be symmetric under $\omega \rightarrow -\omega$, 
because of the symmetry properties of $T_i$ and the restriction for $m$ Eq.~(\ref{eq:Cauchy_omega}) .
Thus, 
\begin{eqnarray}
 \frac{d \omega}{\omega^{m}}\, T_i(\omega, Q^{2})
 \stackrel{\omega \rightarrow -\omega }{ \longrightarrow }
+ \frac{d \omega}{\omega^{m}}\, T_i(\omega, Q^{2})
\, , 
\quad\quad\quad
i = 2,3,L
\, ,
\end{eqnarray}
and, in addition, we have used taken the discontinuity across the branch cut
\begin{eqnarray}
{\rm Im} T_i(\omega,{Q^2})=\frac{1}{2 {\rm i}}
(T_i\left(\omega+{\rm i}\epsilon,Q^{2})-T_i(\omega-{\rm i}\epsilon, Q^{2})\right)
\, , 
\quad\quad\quad
i = 2,3,L
\, .
\end{eqnarray}
Changing variables $\omega \to 1/x$ in Eq.~(\ref{eq:CauchyTmunu}) and using Eq.~(\ref{eq:Cauchy_omega}) 
for the r.h.s. of Eqs.~(\ref{eq:invarTmunu2L}), (\ref{eq:invarTmunu3})  one gets
\begin{eqnarray}
\label{eq:momentsForPr}
\frac{1}{\pi} \int\limits_{0}^{1}{d x}\, x^{n-1}\, {\rm Im} T_{i}(x,{Q^2})= \sum_{j} C_{i,j}\left(n,\frac{Q^2}{\m^2},\a_s\right) 
 A_{{\rm{nucl}}}^{j}\left(n,{\m^2}\right)
\, , 
\quad\quad\quad
i = 2,3,L
\, . 
\end{eqnarray}
Applying the Lorentz projectors~(\ref{eq:projL})--(\ref{eq:proj3}) to the 
optical theorem  Eq.~(\ref{eq:opticaltheorem}) and using Eq.~(\ref{eq:htensor}) for hadronic tensor we obtain
\begin{eqnarray}
\label{eq:optThForPr}
\frac{1}{\pi}{\rm Im} T_{i}(x,Q^{2}) &=& \frac{1}{x}F_i(x,Q^{2})\, , 
\quad\quad 
i=L,2
\, ,
\\
\frac{1}{\pi}{\rm Im} T_{3}(x,Q^{2}) &=& F_3(x,Q^{2})
\, .
\end{eqnarray}
The combination of Eq.~(\ref{eq:momentsForPr}) with  Eq.~(\ref{eq:optThForPr}) 
concludes the derivation of Eqs.~(\ref{eq:F2mellinMR}), (\ref{eq:F3mellinMR}) 
with special emphasis on the transformation properties of $T_{\mu\nu}$ 
under $\omega \rightarrow -\omega$ ($q\rightarrow-q$). 

%
% ---------------------------------------------------------------------

\end{document}